\documentclass[review,jog]{igs}

\usepackage[utf8]{inputenc}
 \usepackage{url}
 \usepackage{amsmath}
\usepackage{empheq}
\usepackage{mathabx}
\usepackage{graphicx}
\usepackage{siunitx}
\usepackage{algorithm}
\usepackage{algpseudocode}
\usepackage{lineno}
\usepackage{amssymb}
\usepackage{enumitem}
\usepackage{amsmath}
\usepackage[nopatch]{microtype}
\usepackage{booktabs}
\usepackage{enumitem}
\usepackage{rotating}
\usepackage{booktabs}
\usepackage[utf8]{inputenc}
\usepackage{enumerate}
\usepackage{subcaption}
\usepackage{pythonhighlight}
\usepackage{epsfig}
\usepackage{amsmath}
\usepackage{mathrsfs}  
\usepackage{amsthm}

\usepackage{graphicx}
\usepackage{float}
\usepackage{bm}
\usepackage{graphicx}
\usepackage{multirow}

\renewcommand{\d}{\mathrm{d}}
\newcommand{\Ki}{\mathrm{k}}

\renewcommand{\d}{\mathrm{d}}

\newcommand{\Flux}{\textbf{q}}
\usepackage{amsfonts}
\usepackage{amsmath}

\makeatletter
\newcommand*{\da@rightarrow}{\mathchar"0\hexnumber@\symAMSa 4B }
\newcommand*{\da@leftarrow}{\mathchar"0\hexnumber@\symAMSa 4C }
\newcommand*{\xdashrightarrow}[2][]{%
  \mathrel{%
    \mathpalette{\da@xarrow{#1}{#2}{}\da@rightarrow{\,}{}}{}%
  }%
}
\newcommand{\xdashleftarrow}[2][]{%
  \mathrel{%
    \mathpalette{\da@xarrow{#1}{#2}\da@leftarrow{}{}{\,}}{}%
  }%
}
\nolinenumbers

\newcommand*{\da@xarrow}[7]{%
  \sbox0{$\ifx#7\scriptstyle\scriptscriptstyle\else\scriptstyle\fi#5#1#6\m@th$}%
  \sbox2{$\ifx#7\scriptstyle\scriptscriptstyle\else\scriptstyle\fi#5#2#6\m@th$}%
  \sbox4{$#7\dabar@\m@th$}%
  \dimen@=\wd0 %
  \ifdim\wd2 >\dimen@
    \dimen@=\wd2 %
  \fi
  \count@=2 %
  \def\da@bars{\dabar@\dabar@}%
  \@whiledim\count@\wd4<\dimen@\do{%
    \advance\count@\@ne
    \expandafter\def\expandafter\da@bars\expandafter{%
      \da@bars
      \dabar@ 
    }%
  }%
  \mathrel{#3}%
  \mathrel{%
    \mathop{\da@bars}\limits
    \ifx\\#1\\%
    \else
      _{\copy0}%
    \fi
    \ifx\\#2\\%
    \else
      ^{\copy2}%
    \fi
  }%
  \mathrel{#4}%
}
\makeatother

\renewcommand{\d}{\textrm{d}}

\usepackage{amsmath}
\usepackage[nopatch]{microtype}
\usepackage{booktabs}
\usepackage{xcolor}

\newcommand{\porosity}{\phi}

\jourvolume{XX}
\jourissue{X}
\jourpubyear{20XX}

\begin{document}


\title[HydroFirn: A numerical model for firn hydrology]{HydroFirn: A numerical model for large-scale multidimensional firn hydrology}
\author[Shadab and others]{Mohammad Afzal Shadab$^{1,2,\dagger}$, Surendra Adhikari$^{3}$, C.~Max Stevens$^{4,5}$, {\AA}sa K. Rennermalm$^{6}$, Jing Xiao$^{6}$, Marc A. Hesse$^{7,8}$, and Reed M. Maxwell$^{1,2,9}$}

\affiliation{
$^{1}$ Department of Civil and Environmental Engineering, Princeton University, Princeton NJ 08544 \\
$^{2}$ Integrated GroundWater Modeling Center, Princeton University, Princeton NJ 08544 \\
$^{3}$ Jet Propulsion Laboratory, California Institute of Technology, Pasadena CA 91109 \\
$^{4}$ Earth System Science Interdisciplinary Center, University of Maryland, College Park MD 20740 \\
$^{5}$ Cryospheric Sciences Laboratory, NASA Goddard Space Flight Center, Greenbelt, MD 20771 \\
$^{6}$ Department of Geography, Rutgers, The State University of New Jersey, Piscataway, NJ 08854 \\
$^{7}$ Department of Earth and Planetary Sciences, University of Texas at Austin, Austin TX 78712 \\
$^{8}$ Oden Institute for Computational Engineering and Sciences, University of Texas at Austin, Austin, TX 78712\\
$^{9}$ High Meadows Environmental Institute, Princeton University, Princeton NJ 08544 \\
Correspondence: Mohammad Afzal Shadab
\email{mashadab@princeton.edu}}

\begin{frontmatter}

\maketitle

\begin{abstract}
Observations show the multidimensional dynamics of meltwater and distribution of ice layers in the firn on the Greenland Ice Sheet. However, state-of-the-art large-scale models for firn hydrology are essentially one-dimensional, limiting their ability to explain observed datasets and failing to reduce uncertainty in surface mass balance and sea-level rise estimates. Here, we present a large-scale, multidimensional, multiphase, and thermomechanical model to simulate firn hydrology. The model is highly efficient due to a novel algorithm in which an extra equation for pressure is solved only in saturated regions. Furthermore, the model can apply spatially heterogeneous boundary conditions to the unsaturated-saturated domain and allows for the dynamic formation of fully impermeable ice layers. The numerical results show excellent comparisons against analytic solutions to one- and two-dimensional problems that involve coupled unsaturated-saturated flows, thermodynamics, and phase change. We further apply the model to investigate field data from southwest Greenland and find that lateral heterogeneities strongly influence the depth of melt percolation and ice layer formation. Improved understanding of these local, multidimensional processes will provide physics-based constraints on firn densification, reduce uncertainty in converting altimetric elevation change to mass change, and improve estimates of freshwater fluxes to the ocean under a warming climate.


\end{abstract}

\end{frontmatter}

\section{  Introduction}\label{sec1}

Rapid mass loss from glaciers and ice sheets is a major contributor to contemporary sea-level rise \citep{Velicogna2020,imbie2018mass,mouginot2019forty,zemp2019global,firn2024firn}. Over the past several decades, surface melting across polar ice sheets has intensified and expanded inland to higher elevations \citep{VanAngelen2013,bell2018antarctic,horlings2022expansion}. On the Greenland Ice Sheet, surface melt and associated runoff account for a substantial fraction of total mass loss \citep{van2009partitioning,Machguth2016}. In regions covered by sintered and compacted snow, called firn, meltwater can infiltrate below the surface and refreeze, forming ice layers within the firn column \citep{harper2012greenland,tc-9-1203-2015}. {The processes such as meltwater freezing and storage may delay runoff.} Under continued atmospheric warming, however, progressive densification of firn is expected to reduce pore space as well as cold content and weaken this melt buffering mechanism, thereby increasing freshwater discharge to the ocean \citep{Pfeffer1991,VanAngelen2013,Noel2017,vandecrux2020}.

Dense and less permeable ice occurs within firn across a broad spectrum of spatial scales, ranging from thin, discontinuous ice lenses to thick, laterally extensive layers and slabs \citep{culberg2021extreme,tedstone2025concurrent}. Observations from firn cores, ground-based and airborne radar, and satellite measurements demonstrate that such features are widespread across Greenland \citep{van2009partitioning,Machguth2016,macferrin2019rapid,samimi2020meltwater,culberg2021extreme,Jullien2023}, as well as in the Canadian Arctic \citep{rutishauser2016characterizing,chan2022spatial,Gascon2013} and Antarctica \citep{jiahong1998glaciological,christoffersen2003response,kaczmarska2006ice,alley2018quantifying}. As these ice layers thicken and connect laterally, they can inhibit vertical percolation, promote the formation of perched water bodies, and redirect meltwater horizontally within the firn \citep{tc-9-1203-2015,Machguth2016,macferrin2019rapid,
Jullien2023}. Such reorganization of subsurface flow pathways enhances the likelihood of runoff generation and can accelerate ice-sheet mass loss \citep{harper2012greenland,Machguth2016}.

Despite their significance, the processes responsible for the initiation, growth, and spatial organization of impermeable ice layers are not yet well understood. {Ice layer formation requires freezing to become localized over a narrow depth interval, which may arise through meltwater ponding above stratigraphic contrasts \citep{Marsh1984a,Pfeffer1998,wever2016simulating,Humphrey2021} or through rapid freezing of infiltrating meltwater when over-steepened thermal gradients ahead of infiltration fronts relax as liquid water content declines \citep{shadab2024mechanism}.} In practice, low-permeability layers likely emerge through repeated infiltration and refreezing events that progressively reduce porosity toward pore close-off \citep{shadab2024mechanism}. As permeability decreases, meltwater fluxes can exceed the local hydraulic conductivity of the firn, leading to perching, saturation, and lateral flow. Accurately representing these transitions, from unsaturated to fully saturated conditions, is therefore essential for predicting the complete firn hydrology from the unsaturated to the saturation region and estimating the partitioning of meltwater into liquid storage, refreezing, and runoff.

Simulating meltwater infiltration in firn remains challenging, and current models show substantial disagreement in their predictions \citep{Stevens2020,vandecrux2020,firn2024firn}. In particular, existing models struggle to reproduce the formation of ice layers, capture lateral flow, and represent the interaction of meltwater with ice lenses and slabs \citep{firn2024firn}. Many widely used firn hydrologic models rely on vertically one-dimensional representations due to computational constraints \citep[see][for a summary]{steger2017firn,vandecrux2020,firn2024firn}, and in many cases, employ bucket-type schemes to approximate percolation and refreezing \citep{coleou1998irreducible,Bartelt2002,Ligtenberg2011,KuipersMunneke2015b,Vionnet2012,Verjans2019}. More physics-based formulations apply Darcy-type flow laws to unsaturated firn, either through kinematic wave approaches \citep{colbeck1974water,Jordan1991,Clark2017,shadab2025unified} or Richards' equation \citep{Illangasekare1990,Wever2014,meyer2017continuum,moure2023thermodynamic,shadab2024mechanism}. Multidimensional firn hydrologic models have been developed that also include preferential flow \citep{Illangasekare1990,Hirashima2014,leroux2020simulation,moure2023thermodynamic}, but they typically require very high spatial resolution, which limits domain size and makes them computationally prohibitive for large-scale climate applications \citep{wever2016simulating,firn2024firn}. Although these frameworks capture key aspects of meltwater transport, they are generally restricted to unsaturated conditions or simplified geometries and do not resolve the formation of saturated regions and associated pressure-driven flow in multiple dimensions nor do they handle the formation of impermeable ice layers. Consequently, a unified multidimensional framework capable of dynamically simulating the coupled evolution of unsaturated and saturated flow, along with the formation of impermeable ice layers at large spatial scales, remains lacking.

The strong coupling between mass transport, energy balance, and phase change introduces nonlinearities that complicate both numerical implementation and model evaluation. As a result, existing firn hydrology models exhibit a large spread in simulated meltwater percolation and ice layer properties, including ice layer depth, thickness, and lateral continuity \citep{vandecrux2020}. Analytic solutions to idealized problems provide an essential benchmark for testing numerical schemes and isolating the effects of discretization and coupling strategies \citep{colbeck1978physical,Clark2017,shadab2025unified}. Kinematic wave models are particularly valuable in this regard because they retain the dominant nonlinear physics of meltwater transport while remaining analytically tractable under simplified conditions. Classical kinematic theory, however, breaks down once local saturation occurs, as flow of liquid water transitions from being governed by a hyperbolic partial differential equation in the unsaturated region (gravity-driven) to being governed by an elliptic PDE (hydraulic pressure-driven) in saturated regions \citep{shadab2022analysis,shadab2024hyperbolic,shadab2025unified}.

\citet{shadab2024hyperbolic} have shown that in the absence of capillary forces, hybrid 1D-3D hydrologic models can be developed efficiently for large-scale problems in multidimensions, with one-dimensional gravity drainage in the unsaturated medium and multidimensional hydraulic pressure gradient driven dynamics in saturated regions. Building on this foundation, the present study introduces a large-scale, multidimensional, multiphase thermo-hydrologic model called HydroFirn that solves coupled mass and energy transport, phase change, and allows transitions between unsaturated and saturated flow, as well as the formation of ice layers. The continuum model is presented in Section \ref{sec2:model-formulation}, and its numerical implementation using the conditionally implicit pressure and explicit enthalpy and composition (CIMPEC) algorithm is given in Section \ref{sec3:numerical-model}. The model is verified against analytic solutions for challenging benchmark problems in one and two dimensions in Section \ref{sec4:verification}. In Section \ref{sec5:Dye2_study}, the simulator is applied to model high-resolution field observations from the DYE-2 site in southwest Greenland to investigate how lateral heterogeneity and surface forcing shape subsurface meltwater pathways and ice layer formation. By resolving the multidimensional meltwater dynamics and the processes leading to the formation of saturated regions and ice layers, this model will help advance a physics-based understanding of firn densification, surface mass balance, and ice-sheet contributions to sea-level rise.

\section{  Continuum Model Formulation} \label{sec2:model-formulation}

In this section, we first define the conserved quantities, then introduce the governing equations and constitutive models, and finally provide the resulting dimensionless continuum model. The related assumptions will be introduced when required. {The enthalpy-based continuum model formulation was previously developed in one (depth) dimension numerically
in \cite{shadab2024mechanism} and theoretically, in limit of no heat conduction, in \cite{shadab2025unified} to derive analytical solutions based on a unified kinematic wave theory for simple melt infiltration problems involving discontinuities in firn conditions.} Here, we extend this framework to multiple dimensions, incorporate heat conduction, and focus on its numerical implementation for general problems involving firn hydrology.

\subsection{ Conserved Quantities}
\begin{figure*}
    \centering
    \includegraphics[width=\linewidth]{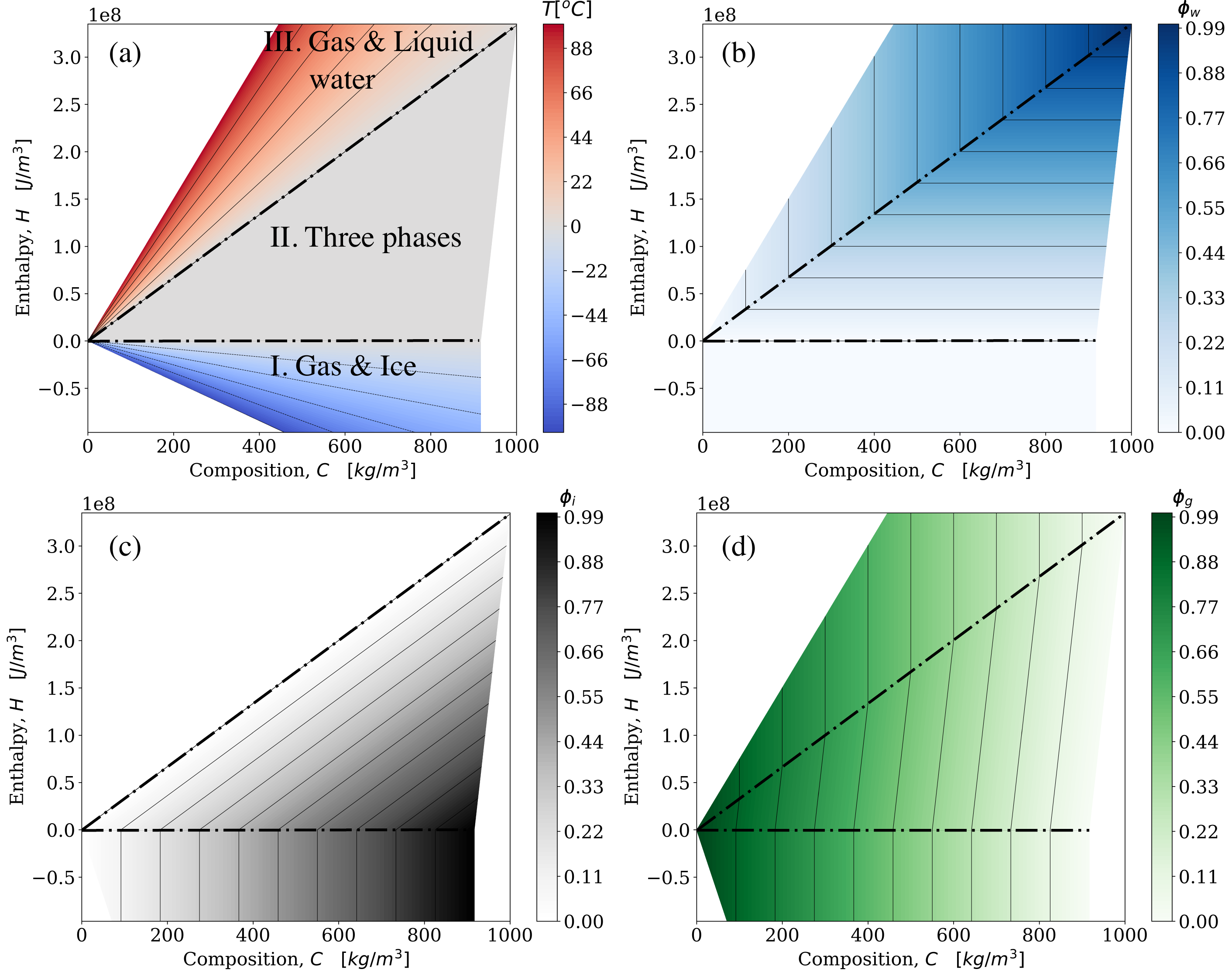}
    \caption{The dependence of temperature and volume fractions on dimensional enthalpy and composition ($C,H$), respectively: (a) temperature and volume fractions of (b) liquid water $\phi_w$ (or liquid water content), (c) ice  $\phi_i$, and (d) gas  $\phi_g$ phases. The contours are restricted to $T\in[-100^\circ\textrm{C},100^\circ\textrm{C}]$ to avoid phase change at boiling as well as to keep the contour levels consistent. Dashed black lines are the level-sets whereas the dashed-dot lines show the boundaries of the regions, i.e., $H=0$ and $H=CL$. The three regions are labeled in panel \textit{a}.}
    \label{fig:combined-variables}
\end{figure*}

Firn is considered a three-phase system comprising liquid water ($w$), ice ($i$), and gas ($g$). These three phases are composed of two components, namely, water (H$_2$O) and air ($\sim$Nitrogen gas, N$_2$). The water component partitions into liquid and solid phases but not into the gas phase. While water vapor plays an important role in firn processes such as grain metamorphism \citep{mcdowell2023firn}, in this work we consider its effects on the physics of meltwater percolation and refreezing negligible. As a result, the air component is confined to the non-reactive gas phase. The conserved variables are the  water composition $C$ (kg/m$^3$), defined as the total mass of water component per unit representative elemental volume (REV) and enthalpy of the system (J/m$^3$), $H$ per unit REV due to phase change involved \citep{Jordan1991,Alexiades1993,aschwanden2012enthalpy,Carnahan2021}. Mathematically, they are defined as

\begin{align} \label{eq:C-working-def}
   C &:= \rho_i \phi_i + \rho_w \phi_w,\\
    H &:= \begin{cases}\rho_i c_{p,i} \phi_i (T-T_m), & {T < T_m} \quad (\textrm{or } H \leq 0) \\  \rho_w \phi_w L,  &{T= T_m} \quad (\textrm{or } 0 < H < CL) \\ \rho_w \phi_w \left(c_{p,w} (T-T_m) + L \right), &{T> T_m} \quad (\textrm{or } H \geq C L) \end{cases}
    ,\label{eq:enthalpy-formulation}
\end{align}
with the constraint $\phi_w + \phi_i \leq 1$ where $\rho_\alpha$ is the density (kg/m$^3$), $\phi_\alpha$ refers to volume fraction of the phase $\alpha \in \{w,i,g \}${, where the mathematical symbol $\in$ means ``belongs to'', and} $T$ is the temperature of the firn (K). The formulation assumes that the water vapor component is negligible in the gas phase and that ice and water phases are pure.
{For simplicity we fix the reference enthalpy at the melting point} to be $H=0$ where the system has no liquid water at the melting temperature, $T=T_m$. Here $c_{p,\alpha}$ is the specific heat capacity at constant pressure (J/kg$\cdot$K) for phase $\alpha$, $T_m$ is the melting temperature (K) and $L$ is the latent heat of fusion of water (J/kg). The density and specific heat capacity of gas are much lower than those of liquid water or ice (see Table \ref{table:1}). We make the simplification that the gas phase contribution to the total enthalpy of the system is negligible. The maximum enthalpy limit for the three-phase region is the product of composition $C$ and the latent heat of fusion $L$, because it is the enthalpy of the fully molten system at the melting point (see Figure \ref{fig:combined-variables}\emph{a}). {The boundaries of the three-phase region, defined by the lines $H=0$ and $H=CL$, are not included in the region $0 < H<CL$ because it strictly refers to the three-phase region.} 
From the above formulation, we classify three regions: region 1 ($H \leq 0$) comprises ice and gas phases, region 2 ($0 < H<CL$) contains all three phases, and region 3 ($H \geq CL$) corresponds to a no-matrix state consisting of only liquid water and gas phases.
The temperature and volume fractions of liquid water, ice, and gas phases can be evaluated from composition, $C$, and enthalpy, $H$, as shown in Figures~\ref{fig:combined-variables}\emph{a-d}, respectively. Next, we formulate the governing equations for this model corresponding to the two conserved variables.

\subsection{  Transport Model}
The conservation equations for water composition and system enthalpy are respectively given as

\begin{align}
    \frac{\partial C}{\partial t} + \nabla \cdot (\textbf{q} \rho_w) &= 0, \quad \forall \textbf{x} \in \Omega  \backslash \partial \Omega,~ t >0,\label{eq:comp-conservation-final}\\
    \frac{\partial H}{\partial t} + \nabla \cdot (\textbf{q} \rho_w \left(c_{p,w} (T-T_m) + L \right) - \overline{\kappa} \nabla T) &= 0, \quad \forall \textbf{x} \in \Omega  \backslash \partial \Omega,~ t >0, \label{eq:enthalpy-conservation}
\end{align}
where $\textbf{x}=(x,y,z)^T$ is the location vector with $z$ being positive downwards, $\textbf{q}$ is the volumetric flux of the water phase (m$^3$/m$^2\cdot$s) relative to the ice phase, $\Omega$ is the domain, and $\partial \Omega $ is the boundary of the domain. {Figure~\ref{fig:potato_diagram} shows a schematic diagram of a variably saturated firn with multiple disconnected saturated regions and impermeable ice layers. A region in the case of firn may represent the entire firn (domain) or an individual control volume (sub-domain). Its boundary is the closed surface that encloses the region and across which total water and enthalpy are exchanged with neighboring regions or the external environment.} The notation $\forall$ means ``for all'' and $\backslash$ denotes set exclusion. Here $\overline{\kappa}$ is the effective thermal conductivity of the firn defined as ${\overline{\kappa}}:=\kappa_i \phi_i^l + \kappa_w\, \phi_w $ with $\kappa_i$ and $\kappa_w$ being the thermal conductivity of the ice and water phases (W/m$\cdot$K) and $l$ being the power-law exponent with $l=1.885$ \citep{yen1981review}.


\subsection{  Constitutive Relations}
The volumetric flux of water relative to ice, $\textbf{q}$, can be written using extended Darcy's law,

\begin{align} \label{eq:darcy-full}
    \textbf{q} = - \frac{\Ki(\porosity) k_{rw}(s_w)}{\mu} (\nabla p - \rho \textbf{g})
\end{align}
where $\Ki$ is the absolute permeability (m$^2$) which is a function of porosity $\porosity$ (i.e., the ratio of void volume to the bulk volume: $\porosity=\phi_w+\phi_g =1-\phi_i$), $p$ is water pressure (Pa), $\mu$ is the viscosity of water (Pa$\cdot$s) and $\textbf{g}$ is the acceleration due to gravity vector (m/s$^2$). The relative permeability for multi-phase flow $k_{rw}$ displays complex hysteresis \citep{blunt2017multiphase}, but here we only consider the simplest case with power law dependence. The relative permeability of the water phase, $k_{rw}$, is a function of the water saturation, $s_w$, which is the ratio of water phase volume to void volume, i.e., $s_w= \phi_w/(1-\phi_i)$.  We assume that the water phase becomes immobile below a certain residual water saturation, $s_{w r}$. Similarly the gas phase becomes immobile below the residual gas saturation, i.e., zero for firn. As a result, the two-phase fluid flow of both gas and water phases is restricted to regions where $s_{wr}<s_w<1$. We will refer to regions with $s_w=1$ as saturated in the remainder of this paper. 

\begin{figure}
    \centering
    \includegraphics[width=0.6\linewidth]{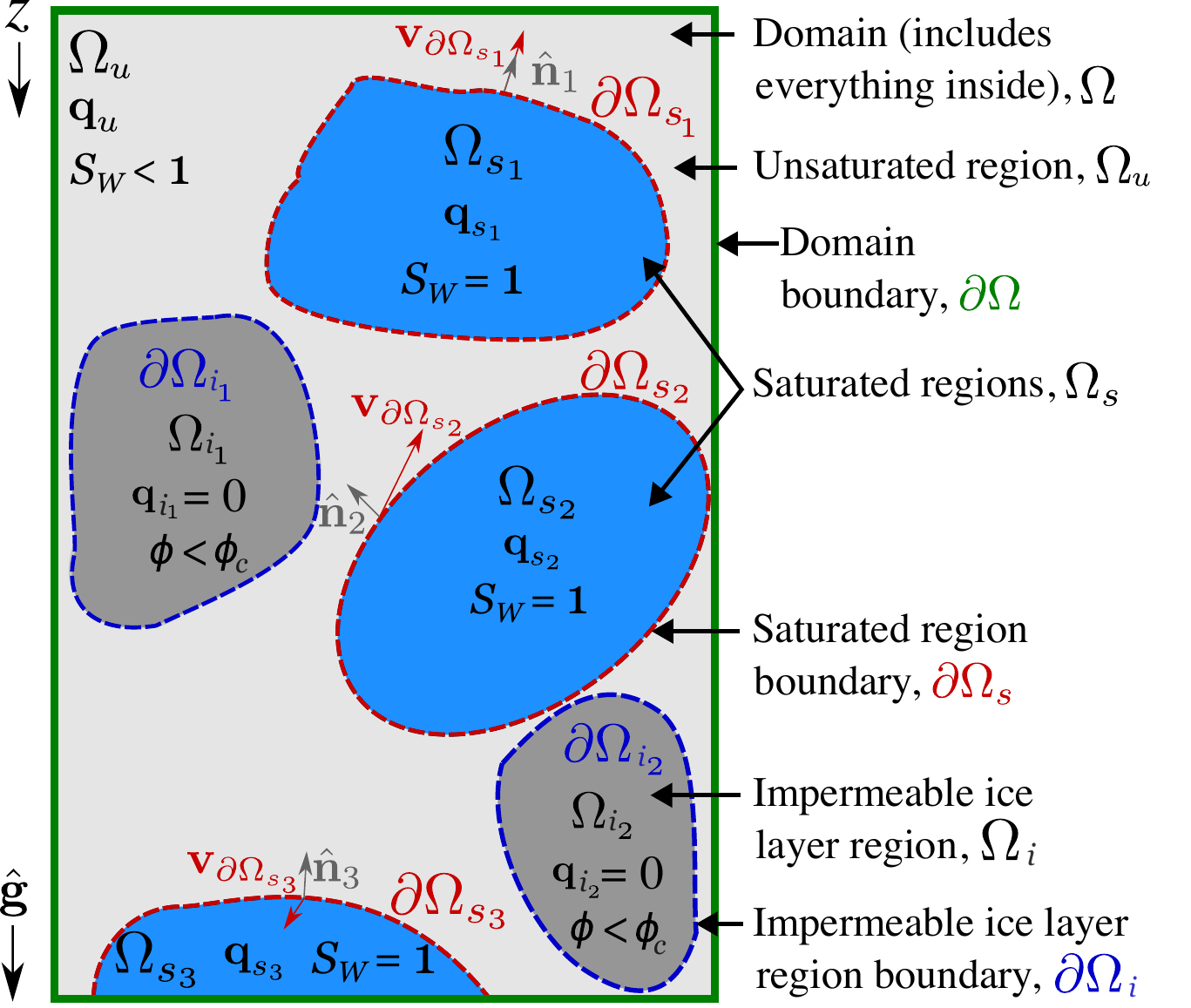}
    \caption{A schematic diagram illustrating variably saturated flow inside firn at the continuum scale with a domain $\Omega$ and its boundary $\partial \Omega$ is denoted with a solid green line. The domain includes unsaturated region $\Omega_u$ with mobile air ($s_w<1$) and three fully saturated subdomains (blue shaded region, $s_w=1$) constituting $\Omega_s \equiv \bigcup_{k=1}^{3}\Omega_{s_k}$, where $\bigcup$ denotes the union operator. Also, for the $k^\textrm{th}$ saturated subdomain ($k\in\{1,2,3 \}$), $\hat{\textbf{n}}_k$ is the outward normal vector to the corresponding saturated-unsaturated region boundary $\partial \Omega_{s_k}$ (red dashed line), $\textbf{v}_{{\partial \Omega}_{s_k}}$ is the velocity of the interface and $\textbf{q}_{s_k}$ is the spatially varying volumetric flux of water phase, at a specific location. Furthermore, $\Omega_{i} \equiv \bigcup_{l=1}^{2} \Omega_{i_l}$ are the regions of impermeable ice layers where the porosity $\phi$ of the representative elemental volume is smaller than the cut-off porosity $\phi_c$ and the permeability goes to zero. The ice layers are bounded by the boundary $\partial \Omega_{i_l}$ shown by dashed blue lines. Furthermore, $\hat{\textbf{g}}$ is the direction of the gravity here shown to be vertically downwards, i.e., in the direction of increasing depth variable $z$. The volumetric flux of liquid water within the unsaturated region is given by $\textbf{q}_u$, in the saturated region is given by $\textbf{q}_s$, and within ice layer is $\textbf{q}_i=0$.}
    \label{fig:potato_diagram}
\end{figure}

{Next we discuss the observed values of residual saturation and cut-off porosity before assigning their values in the model.} The residual water saturation during drainage has been estimated to be approximately 0.07 {m$^3$/m$^3$} from lysimeter \citep{colbeck1976analysis} and calorimeter \citep{coleou1998irreducible} techniques. However, as the ice is water-wet with a near-zero contact angle at the ice-water-air interface \citep{knight1971experiments}, the residual water saturation during saturation rise (imbibition) is zero due to hysteresis in the relative permeability-capillary pressure curve \citep{carlson1981simulation,blunt2017multiphase}. {Therefore, a fixed value of residual water saturation $s_{wr}$ is not accurate for both imbibition and drainage processes.} Furthermore, we refer to the region where the porosity $\phi$ becomes less than or equal to the cut-off porosity {$\phi_c$} as the impermeable ice layer (gray region bounded by blue dashed lines in Figure~\ref{fig:potato_diagram}). The cut-off porosity is typically {observed to be $\phi_c= 0.094$} corresponding to a density of 830 kg/m$^3$ \citep{cuffey2010physics}.

Next, we assume the problem is gravity dominated in unsaturated regions \citep{colbeck1972theory}, such that the spatial variability in the capillary pressure (i.e., the difference between the water and air pressure) is negligible at the problem length scales \cite[e.g., ][]{Smith1983,shadab2022analysis,shadab2024hyperbolic,shadab2025unified}. As a result, the pressure of the water phase in the unsaturated regions becomes a constant, equal to the reference gas pressure, i.e., $p=0$ \citep{colbeck1972theory,shadab2022analysis}. Applying this assumption to Equation \eqref{eq:darcy-full} eliminates the pressure term in unsaturated regions. The volumetric flux of water, $\textbf{q}$, then takes the gravity-driven form in unsaturated regions given by

\begin{align} \label{eq:darcy-law-simplified}
      \textbf{q} =  \frac{\Ki(\porosity) k_{rw}(s_w)}{\mu}  \rho \textbf{g} \quad \text{for } s_w<1 \text{ and } \phi > \phi_c.
\end{align}
The intrinsic permeability of firn (m$^2$) $\Ki$ and the relative permeability of water $k_{rw}$ are assumed to be power laws \citep{kozeny1927uber,carman1937fluid,brooks1964hydraulic,bear2013dynamics,meyer2017continuum} defined as

\begin{align}
\Ki(\porosity) &=\Ki_0 \porosity^m = \Ki_0 (1-\phi_i)^m , \label{eq:abs-perm-def} \\
k_{rw}(s_w) &= k^0_{rw} s_w^n = k^0_{rw} \left( \frac{\phi_w}{1-\phi_i}\right)^n , \label{eq:rel-perm-def}
\end{align}
where $\Ki_0$ is a permeability constant (m$^2$), and $k^0_{rw}$ is the endpoint relative permeability of the water phase. {Because we are not implementing hysteresis in the relative permeabilities here, we assume that the residual saturation of water phase is zero, i.e., $s_{wr}=0$.} A zero residual saturation will provide accurate speeds for the wetting fronts moving into dry firn due to hysteresis in the relative permeability. {We use a cut-off porosity of $\phi_c=0.094$ corresponding to a density of 830 kg/m$^3$ in this model.} Inserting Equations \eqref{eq:abs-perm-def} and \eqref{eq:rel-perm-def} into Equation \eqref{eq:darcy-law-simplified} finally gives
\begin{align}\label{eq:darcy-law-final}
      \textbf{q}(\phi_i,\phi_w) = \begin{cases}
      \textbf{0}, &H\leq 0 \textrm{ or } \phi \leq \phi_c  \\ \textbf{q}_u = K_h (1-\phi_i)^m \left( \frac{\phi_w}{1-\phi_i}\right)^n \hat{\textbf{g}}, & 0<H<CL \textrm{ and } {\phi_w<(1-\phi_i)} \\ \textbf{q}_s =  - K_h (1-\phi_i)^m \nabla h, & 0<H<CL \textrm{ and } {\phi_w=(1-\phi_i)} 
      \end{cases},
\end{align}

where the acceleration due to gravity vector is $\textbf{g}=g\hat{\textbf{g}}$, with $\hat{\textbf{g}}$ being the unit vector in the direction of gravity, and $K_h = \frac{\Ki_0 k^0_{rw}}{\mu}\rho g$ is a constant. The volumetric flux of liquid water is $\textbf{q}=0$ when there is no liquid water $H\leq 0$ or when an impermeable ice layer has formed with $\phi \leq \phi_c$, leading to no advection and only the transport of enthalpy via heat conduction. When the medium saturates completely, the composition balance \eqref{eq:comp-conservation-final} is used with $\partial C/\partial t = 0$ and Darcy's law \eqref{eq:darcy-full} is applied to the medium that saturates locally ($\phi_w = 1-\phi_i$) and hydraulic gradients couple the flow in all directions across the saturated region, where $h = p_w/(\rho_w g) - \int_0^\textbf{x}\hat{\textbf{g}} \cdot \d \textbf{x}$ is the hydraulic head (units in m) \citep[see][]{shadab2024hyperbolic}. The total mass balance of liquid water then limits to the elliptic equation for incompressible saturated flow

\begin{align}\label{eqn:laplace}
    -\nabla\cdot\left(K_h (1-\phi_i)^m\nabla h\right) &= 0  \quad \forall \textbf{x} \in \Omega_s(t) \textbackslash \partial\Omega_s(t).
\end{align}
  Solving variably saturated flow problems in the gravity-driven limit requires a dynamic coupling between the hyperbolic PDE \eqref{eq:comp-conservation-final} for unsaturated regions and the elliptic PDE \eqref{eqn:laplace} for saturated regions. Although the elliptic PDE \eqref{eqn:laplace} itself is not time dependent, the saturated domain, $\Omega_s(t)$, changes with time due to its interaction with the unsaturated region. We refer to the interface between the saturated and unsaturated regions simply as the interface and denote it as $\partial\Omega_s(t)$ which may evolve with time. There can be multiple saturated regions that can dynamically form and evolve and interact with each other (see Figure \ref{fig:potato_diagram}). The regions containing impermeable ice layers $\Omega_i$ are not considered a part of saturated region $\Omega_s$. The advective flux $\textbf{q}$ is set to 0 inside the ice layer region $\Omega_i$, including on the boundary of the impermeable ice layer region $\partial \Omega_i$, which may also intersect with the boundary of the saturated region $\partial \Omega_s$, i.e., $\partial \Omega_s \cap \partial \Omega_i $.

{Since the pressure in the unsaturated region is always determined by the gas phase and is hence set to zero \citep{szymkiewicz2013mathematical,lie2019introduction,shadab2022analysis,shadab2024hyperbolic,shadab2025unified}}, the hydraulic head boundary condition along the interface is simply

\begin{align}\label{eq:pressure_bnd_cond}
    h=- \int_\textbf{0}^\textbf{x}\hat{\textbf{g}} \cdot \d \textbf{x} \quad \mathrm{on}\quad \mathbf{x}\in\partial\Omega_s(t),
\end{align}
where $\textbf{0}$ is the location vector of the origin, $(0,0,0)^T$. For example, $h=-z$ within all unsaturated cells when gravity is directed vertically downwards, i.e., $\hat{\textbf{g}}=(0,0,1)^T$. The {multidimensional} velocity of the interface, $\mathbf{v}_{\partial\Omega_s}$, can be determined by the discrete balance of composition (Equation~\ref{eq:comp-conservation-final}) across the interface as

\begin{align}\label{eqn:vel_int}
    \mathbf{v}_{\partial\Omega_s} = \frac{(\mathbf{q}_u-\mathbf{q}_s)\cdot\hat{\mathbf{n}}}{\phi_w-(1-\phi_i) }\mathbf{\hat{n}},
\end{align}
where $\mathbf{\hat{n}}$ is the outward unit normal of the interface. The fluxes along the interface are $\mathbf{q}_u$, $\mathbf{q}_s$, and $\mathbf{q}_i$ in the unsaturated, saturated, and impermeable ice layer regions, respectively {(see Figure \ref{fig:potato_diagram} for example)}. {Further, the volume fraction of water or liquid water content $\phi_w$ is evaluated at the unsaturated side of the interface.} The saturated domain boundary, $\partial\Omega_s$, evolves according to this interface velocity $\mathbf{v}_{\partial\Omega_s}$.

Due to gravity-driven drainage in the unsaturated region, hydraulic pressure gradients in the saturated region, and phase change in cold regions, the water saturation can evolve in both saturated and unsaturated domains. As such, we are
simply evolving the composition \eqref{eq:comp-conservation-final-summary} and enthalpy \eqref{eq:enthalpy-conservation-summary}, but we evaluate the fluxes differently in the saturated, unsaturated, and impermeable ice layer regions. This avoids the explicit tracking of the interfaces, and the mathematical model can be summarized as

\begin{empheq}[box=\fbox]{align}
        &\frac{\partial C}{\partial t} + \nabla \cdot (\textbf{q} \rho_w) = 0 \quad \forall \textbf{x} \in \Omega  \backslash \partial \Omega,~ t>0, \label{eq:comp-conservation-final-summary}\\
    &\frac{\partial H}{\partial t} + \nabla \cdot (\textbf{q} \rho_w \left(c_{p,w} (T-T_m) + L \right) - \overline{\kappa} \nabla T) = 0 
 \quad \forall \textbf{x} \in \Omega  \backslash \partial \Omega,~ t >0, \label{eq:enthalpy-conservation-summary} \\
&\hspace{1cm}\text{with } \textbf{q}(\phi_i,\phi_w) = \begin{cases}\textbf{q}_u=K_h (1-\phi_i)^m \left( \frac{\phi_w}{1-\phi_i}\right)^n\hat{\textbf{g}} \quad \forall \textbf{x}\in \Omega \backslash (\Omega_s \cup \Omega_i),\\ \textbf{q}_s=-K_h (1-\phi_i)^m  \nabla h  \quad \hspace{1cm} \forall \textbf{x}\in \Omega_s \text{ where } \phi_w= 1-\phi_i , \\
\textbf{q}_i=0 \hspace{4.2cm} \forall \textbf{x}\in \Omega_i \text{ where } \phi_i > 1- \phi_c,  \end{cases}  \label{eq:final-flux}\\
    &\hspace{1cm}\text{and } -\nabla\cdot\left(K_h (1-\phi_i)^m \nabla h\right) = 0  \quad \forall \textbf{x} \in \Omega_s(t) \textbackslash \partial\Omega_s(t), \label{eqn:laplace-final} \\
    &\hspace{1cm}\textrm{subject to }       h=- \int_\textbf{0}^\textbf{x}\hat{\textbf{g}} \cdot \d \textbf{x} \quad \forall \mathbf{x}\in\partial\Omega_s(t) \label{eq:pressure_bnd_cond-final}.
\end{empheq}
The conserved variables composition $C$ and enthalpy $H$ directly relate to physical variables such as temperature $T$ and volume fractions of ice, water, and air phases (see Equations~\ref{eq:C-working-def}-\ref{eq:enthalpy-formulation} and Figure~\ref{fig:combined-variables}).
This mathematical model (Equations~\ref{eq:comp-conservation-final-summary}-\ref{eq:pressure_bnd_cond-final}) for coupled unsaturated-saturated region hydrology with thermodynamics and phase change in the limit of negligible capillary forces requires the dynamic coupling of hyperbolic and elliptic subdomains for composition balance with evolving interfaces and includes the formation and evolution of impermeable ice layers. Below we develop an efficient numerical algorithm that addresses the unique nature of this model.

\begin{table}
\caption{A summary of simplified thermodynamic properties as well as flow properties of water in porous ice used in present work. The value for the absolute permeability coefficient is taken from \cite{meyer2017continuum} and thermal conductivity-ice volume fraction power law exponent is taken from \cite{yen1981review}.}
\centering
\begin{tabular}{lccc}
\toprule
 Parameter & & Value & Units   \\
 \midrule
 Density of liquid water & $\rho_w$& 1000 & kg/m$^3$ \\
 Density of ice& $\rho_i$& 917 & kg/m$^3$ \\
 Specific heat of liquid water&$c_{p,w}$& 4186 &  J/(kg K)  \\
  Specific heat of ice & $c_{p,i}$& 2106.1 & J/(kg K)    \\
Thermal conductivity of liquid water &$\kappa_w$&0.606& W/(m K)\\
Thermal conductivity of ice &$\kappa_i$&2.25& W/(m K)\\
Latent heat of fusion of water &$L$&333.55& kJ/kg\\
Melting temperature &$T_m$&273.16& K\\
Thermal diffusivity of ice & $\alpha_T$&1.45 $\cdot \text{10}^{\text{-7}}$& m$^2$/s\\
 Absolute permeability coefficient& $\Ki_0$& 5.56 $\cdot\text{10}^{\text{-11}}$ & m$^2$ \\
 Endpoint relative permeability of liquid water& $k_{rw}^0$& 1.0 &  -  \\
 Porosity-permeability power law exponent& $m$& 3.0 &  -  \\
  Relative permeability - saturation power law exponent& $n$& 2.0 &  -  \\
  Acceleration due to gravity & $g$& 9.81 &m/s$^2$    \\
Dynamic viscosity of liquid water & $\mu_w$&$\text{10}^{\text{-3}}$&Pa s\\
Coefficient of hydraulic conductivity& $K_h$& 5 $\cdot 
\text{10}^{\text{-4}}$ &m/s\\
Cut-off or close-off porosity & $\phi_c$&0.094&$\text{m}^{\text{3}}/\text{m}^{\text{3}}$\\
Thermal conductivity-ice volume fraction power law exponent & $l$&1.885&-\\
\bottomrule
\end{tabular}\label{table:1}
\end{table}

\begin{algorithm}
\caption{Conditionally implicit pressure, explicit enthalpy and composition (CIMPEC) algorithm}\label{algo:hyperbolic-elliptic-PDE-solver}
\begin{algorithmic}[1]
  \While{$t < T$} \Comment{Time loop; {superscript $i$ or $i+1$ shows the timestep}}
\State Calculate \emph{gravity drainage flux} $\Flux_u$ ($=\Flux^i$) \eqref{eq:final-flux} at all cell faces in $\Omega$
\State Flag all the \emph{saturated} cells $\Omega_s$ where $s_w>s_{w,T}$ (threshold saturation) \Comment{$s_{w,T}\lessapprox 1$}
\State Flag all the \emph{impermeable ice layer} cells $\Omega_i$ where $\phi < \phi_c$ (cut-off porosity)
\If{$\Omega_s \neq \{  \}$} \Comment{Saturated region flux evaluation loop}
    \State Set Dirichlet boundary condition \eqref{eq:pressure_bnd_cond-final} on head, $h = - \int_\textbf{0}^\textbf{x}\hat{\textbf{g}} \cdot \d \textbf{x}$, in all unsaturated cells
    \State Set head, $h = - \int_\textbf{0}^\textbf{x}\hat{\textbf{g}} \cdot \d \textbf{x}$, in saturated cells on the domain boundary ($\partial \Omega_s \cap \partial \Omega$) to enable outflow
    \State Solve Laplace-type equation \eqref{eqn:laplace-final} \emph{implicitly} for head $h$ 
    \State Calculate saturated region's face fluxes at all cell faces in $\Omega_s$ using Darcy's law \eqref{eq:final-flux}, i.e., $\Flux_s=-K\nabla h$
    \State Substitute the saturated cells' face fluxes in $\Flux^i$ considering the front motion criteria (\ref{eqn:vel_int}) on $\partial \Omega_s$ \State \Comment{Domain boundaries which are saturated ($\partial \Omega \cap \partial \Omega_s$) only utilize saturated flux unless a boundary condition is specified}
\EndIf
\State Substitute the fluxes $\Flux^i = 0$ for faces in impermeable ice layers in $\Omega_i$ including on boundaries $\partial \Omega_i$
\State Calculate the time step $\Delta t^i$ from CFL condition
\State Update the composition $C^{i+1}$ \emph{explicitly} from Equation \eqref{eq:comp-conservation-final-summary}
\State Update the enthalpy $H^{i+1}$ \emph{explicitly} from Equation \eqref{eq:enthalpy-conservation-summary}
\State Update time counter $t^{i+1}=t^i+\Delta t^i$
\State Calculate state variables such as volume fractions ($\phi_i,\phi_w,\phi_g$) and temperature $T$ at time $t^{i+1}$
  \EndWhile
\end{algorithmic}
\end{algorithm}

\section{ Numerical model}\label{sec3:numerical-model}
We use a conservative finite difference framework on a standard Cartesian, staggered grid to solve the governing Equations \eqref{eq:comp-conservation-final-summary} and \eqref{eq:enthalpy-conservation-summary}. The volumetric flux of water $\textbf{q}$ is evaluated using Darcy's law (Equation~\ref{eq:final-flux}). In unsaturated regions, the volumetric flux of water is the gravity drainage flux, i.e., the unsaturated hydraulic conductivity. In saturated regions, the elliptic PDE \eqref{eqn:laplace-final} is solved for the hydraulic head on saturated cells, subject to the Dirichlet boundary condition (fixed head given by Equation \ref{eq:pressure_bnd_cond-final}) imposed on unsaturated cells. The numerical treatment of the governing equations and boundary conditions, particularly the Dirichlet boundary condition, utilizes the discrete operators described in \cite{shadab2024hyperbolic}.

\begin{figure}
    \centering
    \includegraphics[width=0.8\linewidth]{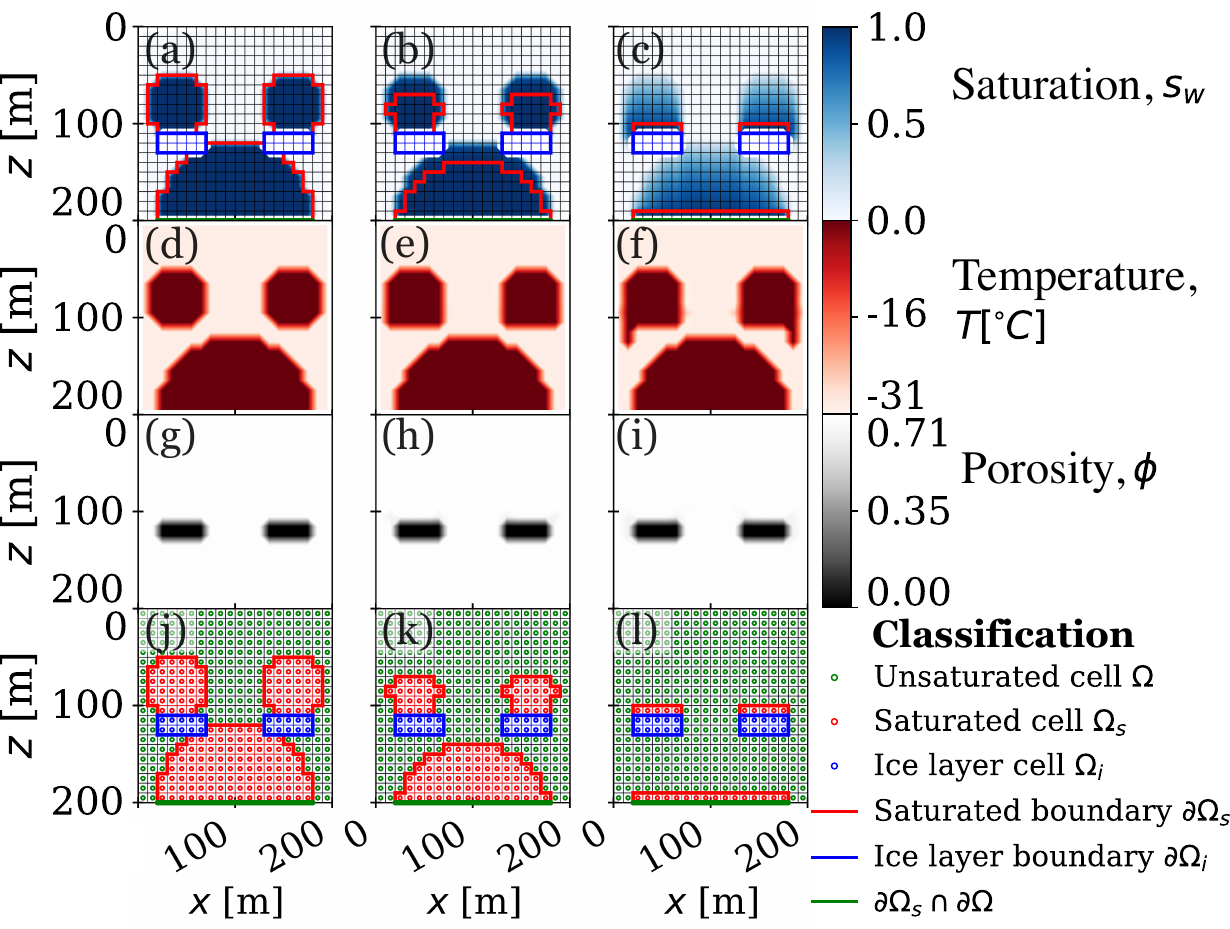}
    \caption{Gravity-dominated drainage of multiple saturated regions across an otherwise dry and cold firn which has two impermeable ice layers. The initial porosity of the entire firn is 0.70 with porosity of initial ice layer being 0.0. The initial temperature of the firn outside the saturated region is -30$^\circ$C whereas inside the saturated region it is 0$^\circ$C. The resulting (a-c) saturation, (d-f) temperature, and (g-i) porosity contours are shown at times $t=$ 0, 1 and 24 hours (left to right), respectively. {(j-l) Their corresponding cell and face classification showing dynamically evolving saturated, unsaturated, and impermeable ice layer cells as well as faces are shown at respective times.} The unit square domain of $200\times 200 $m$^2$ is divided uniformly into $20 \times 20$ grid cells shown by thin lines. The gravity aligns with depth ($+z$) direction, i.e., $\hat{\textbf{g}}=(0,0,1)^T$.}
    \label{fig:HEalgo-schematic}
\end{figure}

Figure \ref{fig:HEalgo-schematic} shows an example of a discretized numerical test problem corresponding to Figure \ref{fig:potato_diagram}. The test shows the gravity drainage of three saturated regions $\Omega_s$ with impermeable ice layers $\Omega_i$ in the domain $\Omega$ which is set to an initial temperature of -30$^\circ$C outside the saturated region. The proposed method efficiently combines the simplicity of solving explicit equations for unsaturated flow (\ref{eq:comp-conservation-final-summary}-\ref{eq:enthalpy-conservation-summary}) with an additional, domain-specific implicit equation \eqref{eqn:laplace-final} to capture the formation and evolution of fully-saturated regions (Figures~\ref{fig:HEalgo-schematic}\emph{a}-\emph{c}).

The saturated subdomains comprising $\Omega_s$ are identified by selecting cells with saturations above a critical threshold, {$s_{w,T}\lessapprox 1$ (e.g., $s_{w,T}=1-10^{-3})$}. Figures~\ref{fig:HEalgo-schematic}\emph{j}-\emph{l} show such cells with red circles at their centers. In the case that any subdomain(s) saturate completely, the flux in the saturated region(s) $\textbf{q}_s$ is then evaluated by solving the elliptic problem \eqref{eqn:laplace-final} subject to Dirichlet boundary conditions \eqref{eq:pressure_bnd_cond-final}. Here we set the heads in all unsaturated cells and eliminate them using the projection approach discussed in \cite{shadab2024hyperbolic}. This approach leads to a reduced system of equations corresponding only to the saturated cells (red circles in Figure~\ref{fig:HEalgo-schematic}\textit{j-l}) and hence efficiently and automatically deals with multiple disconnected saturated regions. For saturated cells on the domain boundary $\partial \Omega_s \cap \partial \Omega $, shown by green lines in Figures \ref{fig:HEalgo-schematic}\emph{j}-\emph{l}, the boundary condition specified on the external boundary must be applied. For outflow boundary conditions, the heads in the corresponding cells must be set to $h=- \int_\textbf{0}^\textbf{x}\hat{\textbf{g}} \cdot \d \textbf{x}$. 
Once the head is evaluated, the flux at the faces inside the saturated region (thin red lines) is evaluated using Darcy's law \eqref{eq:final-flux}, i.e., $\textbf{q}_s=-K \nabla h$. 
The fluxes on the cell faces corresponding to the saturated-unsaturated boundary, $\partial \Omega_s$ shown as a thick red line in Figure \ref{fig:HEalgo-schematic}, are upwinded according to the interface velocity \eqref{eqn:vel_int}. 

Once the fluxes are known, enthalpy $H$ and composition $C$ are evolved explicitly using the time step defined by CFL condition \citep{shadab2024hyperbolic}, the secondary variables $\phi_w$, $\phi$, and $T$ are evaluated from the definitions of composition and enthalpy in Equations (\ref{eq:C-working-def}-\ref{eq:enthalpy-formulation}). This approach is highly efficient, as the extra implicit equation \eqref{eqn:laplace-final} is solved only on the saturated cells. We refer to this algorithmic approach as ``Conditionally Implicit'' Pressure, Explicit Enthalpy, and Composition (CIMPEC) solution approach that is also summarized in Algorithm 1. The resulting numerical simulator is referred to as ``HydroFirn'' in this paper.

\textit{Treatment of impermeable ice layer} ($\phi < \phi_c$): The impermeable ice layers with porosity less than the pore close-off (cut-off) porosity $\phi_c$ are also handled dynamically. The impermeable ice layer regions $\Omega_i$ are excluded from the saturated regions $\Omega_s$ (Figure~\ref{fig:potato_diagram}). This avoids the poor conditioning of the discrete Laplacian matrix that arises from discretizing of pressure equation~\eqref{eqn:laplace-final}. When an impermeable ice layer is formed, the advection of both composition and enthalpy goes to 0 and only the heat can get transported across them via heat conduction. To numerically treat this behavior, we set the volumetric flux of meltwater $\textbf{q}=0$ on all faces corresponding to the ice layer region (all blue lines in Figure~\ref{fig:HEalgo-schematic}{\textit{j-l}}), even if they coincide with the boundary of the saturated region, i.e., $\partial \Omega_i \cap \partial \Omega_s$. In the numerical model, the pore close-off (cut-off) porosity is set to {$\phi_c= 0.094$}, which can be changed.

\textit{Treatment of the surface ablation ($\phi = 1$) {and snow accumulation}}:
{Surface ablation here refers to the complete melting of the firn, whereas snow accumulation is the deposition of fresh snow at the surface.} The ablation of the firn surface by melting is handled by eliminating cells with unit porosity $\phi=1$. The thermal flux boundary condition is then applied to the first cell with non-zero porosity when counted from the top boundary at each horizontal location. {We convert the surface heat flux to an equivalent source term and then apply it to the first cell from top with ice matrix.} {Snow accumulation is denoted by the symbol \textit{a} with units in meter water equivalent per day and is handled by (re-)activating cells above the firn surface. The model integrates the accumulated snow over time until it reaches more than cell height. Once it exceeds the cell height, the model adds fresh, dry snow of density 315 kg/m$^3$ (taken from \cite{vandecrux2020}) at temperature $0\,^\circ$C that corresponds to $C=315$ kg/m$^3$ and $H=0$ J/m$^3$ in one (or more) cell(s) above the surface. $H$ can also be set corresponding to a lower temperature instead of melting temperature if the local air temperature is known.} As such, our grid is Eulerian rather than the Lagrangian grid used in other continuum firn models \cite[e.g.][]{vandecrux2020}. {It is important to note that the published Eulerian models with moving grids such as MeyerHewitt \citep{meyer2017continuum} and DMIHH \citep{langen2017liquid} smooth the porosity structure of the firn, including the newly formed ice layers \citep[see][]{vandecrux2020}. Furthermore, it is not clear how a Lagrangian formulation can be extended to more than one dimension if the ablation or accumulation rates are not uniform. The HydroFirn model with fixed grid and moving surface is able to preserve ice layers and handle accumulation and ablation in multiple dimensions.}
The next section compares the solutions from the HydroFirn model against the analytic solutions for code verification and model validation.

\section{ Model validation}\label{sec4:verification}
In this section, two benchmark tests are performed in the absence of capillarity and heat conduction where analytic solutions are available. These tests involve variably saturated flow along with thermodynamics and phase change.

\subsection{ One-dimensional infiltration leading to formation of a perched aquifer}

{The first validation test considers meltwater infiltration into multilayered firn following a melt event. It combines the well-studied problem of wetting-front propagation in dry firn \citep{colbeck1972theory,gray1996water,Durey2014,meyer2017continuum,shadab2025unified} with the formation and expansion of a perched aquifer into a cold region \citep{shadab2025unified}. Previously, \citet{meyer2017continuum} considered perched aquifer formation in fully temperate firn without phase change. The present test simultaneously evaluates variably saturated flow, heat transport, and phase change by comparing numerical results against analytical solutions.}

\begin{figure}
    \centering
    \includegraphics[width=0.7\linewidth]{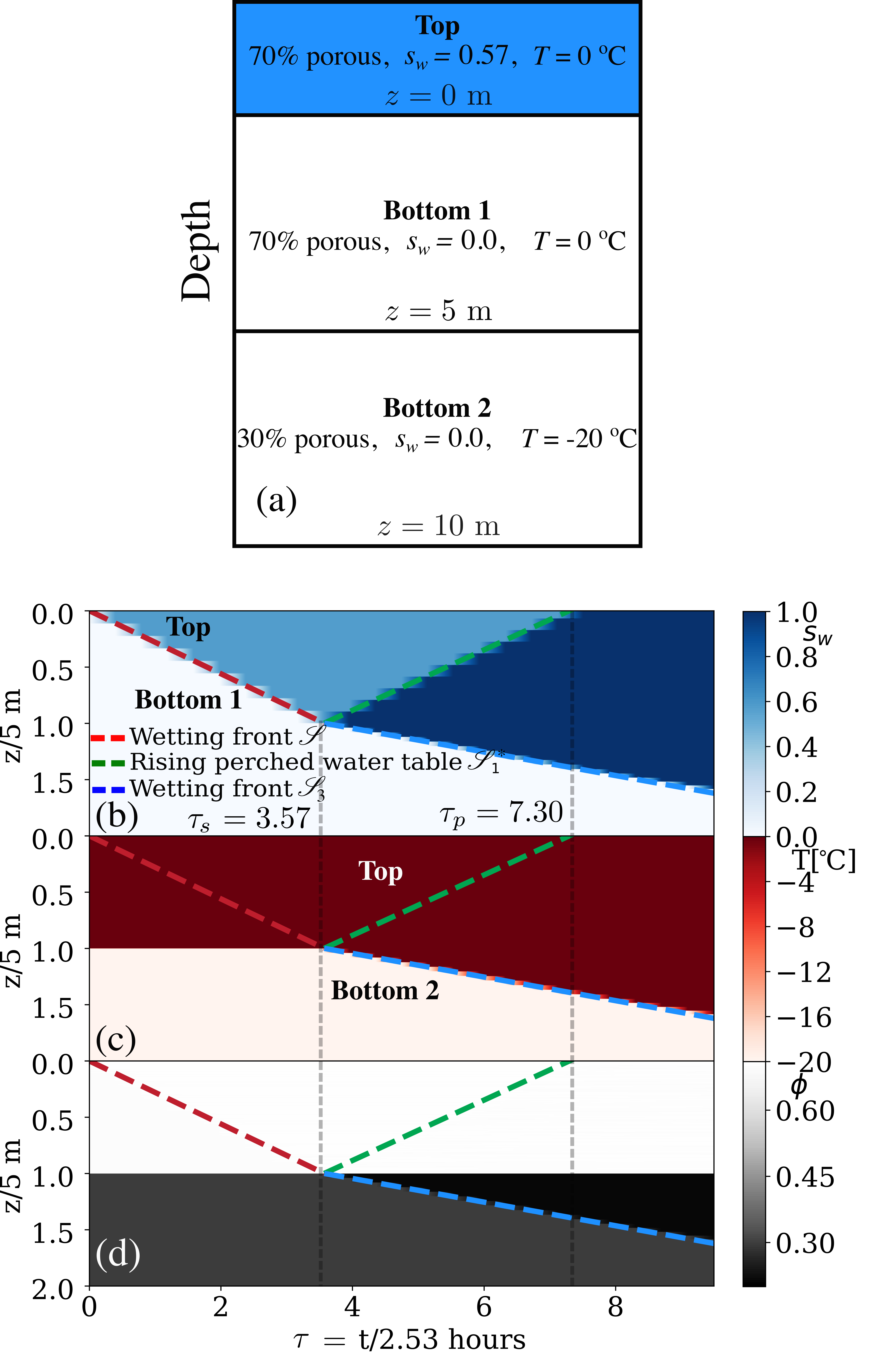}
\caption{Infiltration into a multilayered firn with porosity and temperature decay with depth: {(a) Schematic diagram showing the initial domain with Bottom 1 and 2 layers as well as the boundary (Top) condition.} Contour plots show evolution of the firn (b) saturation $s_w$, (c) temperature $T$, and (d) porosity $\phi$ evaluated by the numerical simulation in absence of heat conduction. Here all dashed lines show analytic solutions computed from the unified kinematic wave theory (UKWT) \citep{shadab2025unified}. The thin, gray dashed {vertical} lines show theoretically calculated dimensionless times of saturation $\tau_s$ and ponding $\tau_p$. The theoretical evolution of the initial wetting front $\mathscr{S}$ (red dashed line) is computed from Case III of UKWT theory whereas the dynamics of saturated region after wetting front $\mathscr{S}$ reaches $z=5$ m shown by blue and green dashed lines is computed by Case XI of the UKWT theory. }\label{fig2:UKWT}
\end{figure}

The firn is initially 70\% porous (density=275 kg/m$^{3}$), dry ($s_w=0$), and at 0$^\circ$C at depths shallower than $z = 5$ m. For depths deeper than 5 m, the firn is initially  30\% porous (density=640 kg/m$^{3}$), dry, and at -20$^\circ$C (Figure~\ref{fig2:UKWT}\textit{a}). {The initial condition is shown in Figure~\ref{fig2:UKWT}\textit{a}}. At time $t=0$ hours, melt is generated at the surface {as a boundary condition} ($z=0$ m), which increases the saturation $s_w$ to 0.57, referred to as ``Top condition'' {and is kept constant for the duration of this simulation (Figure~\ref{fig2:UKWT}\textit{a}). The top boundary condition becomes redundant once the saturated region reaches the surface.} The analytic solution of this problem is given by unified kinematic wave theory proposed in \cite{shadab2025unified} in the limit of negligible capillary forces and negligible effects of heat conduction. Since unified kinematic theory is derived with the assumption of same density (1000 kg/m$^3$) for water and ice phase, the numerical solutions are evaluated with the same assumption for this case. The computational domain $z \in [0,10\textrm{ m}]$ is divided uniformly into 400 cells to keep the cell width at 0.025 m. The boundary condition at the top surface ($z=0$ m) is prescribed to the ``Top condition'' in Figure~\ref{fig2:UKWT}\emph{a}, whereas the bottom boundary condition is not required. 

Initially, a wetting front $\mathscr{S}$ propagates downwards with a constant dimensionless speed (Figures~\ref{fig2:UKWT}\emph{b}-\emph{d}, red dashed line). The time variable is non-dimensionalized as $\tau=t/2.53$ hours where the characteristic time of 2.53 hours comes from dividing the transition depth of 5 m by hydraulic coefficient $K_h$ given in Table~\ref{table:1}. The initial wetting front $\mathscr{S}$ reaches the transition depth of $z=5$ m at $\tau=\tau_s=3.57$ (or $t=3.57\cdot 2.53\sim9$ hours), and a saturated region begins to form. Afterwards, the saturated region forms due to large meltwater flux compared to the hydraulic conductivity of the refrozen region formed right below the transition depth, i.e., $z>5$ m. This saturated region expands in both directions bound by a perched water table ${\mathscr{S}^*_1}$ (top of the perched aquifer shown by green dashed line) that rises to the surface and the wetting front ${\mathscr{S}_3}$ (blue dashed line) percolates downward into the cold region. As the wetting front moves downward, some liquid refreezes and  warms the surrounding firn to the melting temperature. Thus, there is a reduction in porosity from 30\% to 21.2\% across the wetting front for $z>5$ m beginning at dimensionless time $\tau>\tau_s=3.57$ (Figure~\ref{fig2:UKWT}\emph{b}). 
Lastly, ponding starts at a time when the rising perched water table reaches the surface, so the dimensionless ponding time $\tau_p$ can be calculated theoretically as $\tau_p=7.30$ ($t\sim18.5$ hours). All of these dimensionless shock speeds and times are computed analytically, and the resulting locations are plotted with dashed lines in Figures~\ref{fig2:UKWT}\emph{b-d}. The numerical solutions from the HydroFirn model in the absence of conduction, shown by contour plots, demonstrate an excellent comparison with the analytic solutions from the unified kinematic wave theory \citep{shadab2025unified}. 

\subsection{ Two-dimensional firn aquifer migration in cold firn}
\begin{figure}
    \centering
    \includegraphics[width=0.7\linewidth]{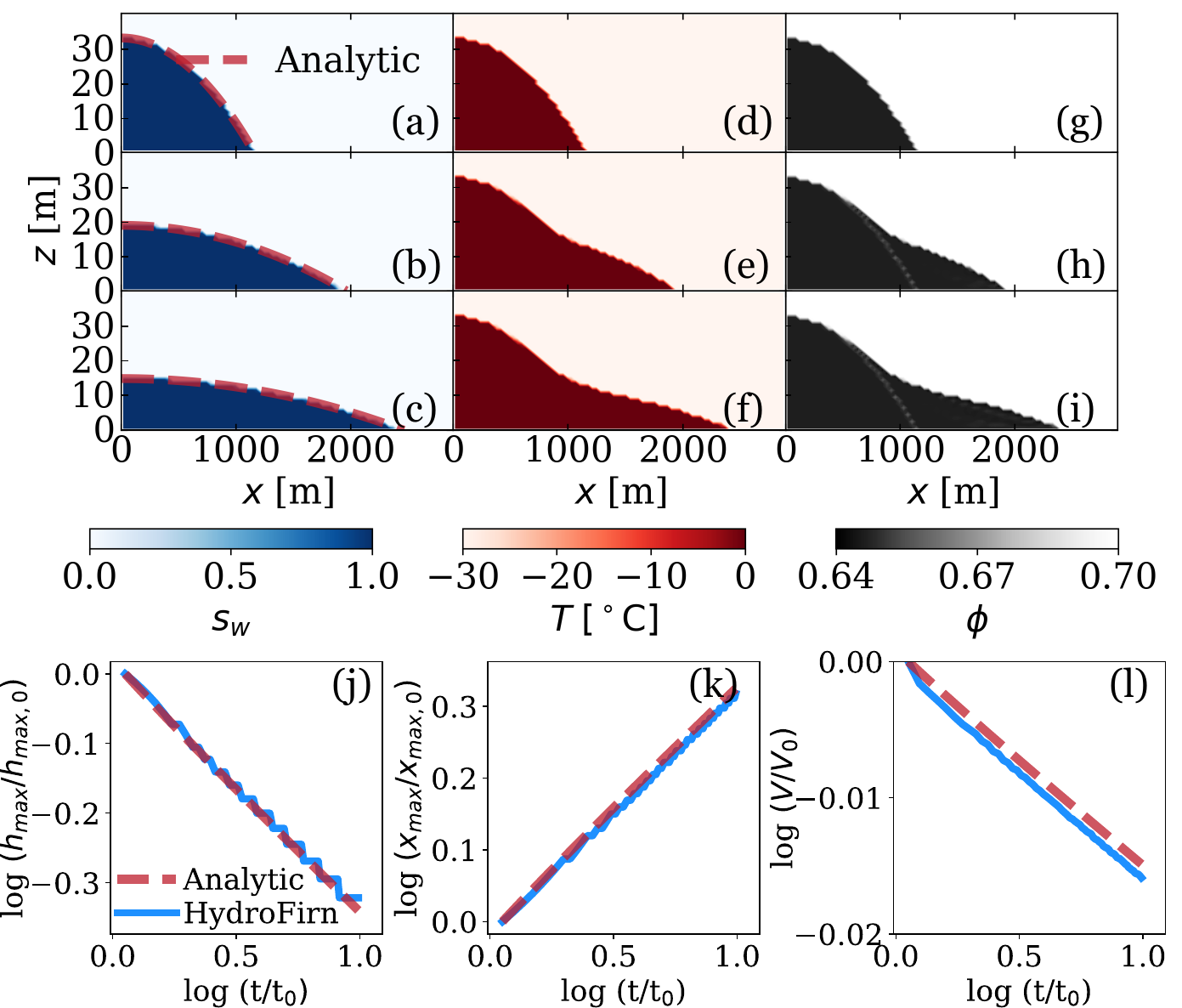}
    \caption{{Expansion of an aquifer in an otherwise uniform, cold firn outside the aquifer with initial temperature $T_0 =-30\,^\circ$C and porosity $\phi_0=0.7$. Solutions of the aquifer height or hydraulic head at (a) $t=1$ year (initial condition), (b) $t=5$ years, and (c) $t=10$ years from theory in cartesian coordinates (Analytic). The contour plots for the same test come from the HydroFirn model {in absence of heat conduction} for (a-c) saturation $s_w$, (d-f) temperature $T$, and (g-i) porosity $\phi$ at $t=$ 1, 5 and 10 years. Evolution of dimensionless (j) maximum height, (k) maximum length, and (l) volume of liquid water in the aquifer, scaled with respect to their initial values at time $t_0=1$ year (subscript $0$ refers to the initial values). HydroFirn model solutions show an excellent comparison against analytical solutions. Logarithms with base 10 ($\log_{10}$) are used in panels \textit{j} through \textit{l}. Supplementary video S1 shows the expansion of firn aquifer from the HydroFirn model in absence of heat conduction (contour plots and solid blue lines) and semi-analytic solutions (red dashed lines).}}
    \label{fig3:firn-aquifer}
\end{figure}

Next, we examine the lateral expansion of a firn aquifer spreading over a horizontal impermeable base within an otherwise cold, porous firn column, a test problem introduced in \cite{shadab2025vertically}. This benchmark helps simultaneously verify multidimensional implementation, saturated region dynamics, heat transport, and phase change. Outside the aquifer, the firn is assumed to be initially homogeneous, with a uniform temperature of $T_0 = -30^\circ$C and an initial porosity of $\phi_0 = 0.7$. Based on energy balance, it leads to a theoretical reduction in porosity of $\Delta \phi = 0.057$, i.e., the porosity should reduce from 0.7 to 0.643 \citep[see][]{shadab2025vertically}. The numerical simulation is initialized using the analytical solution expressed in Cartesian coordinates at $t = 1$~year (Figure~\ref{fig3:firn-aquifer}\textit{a}), corresponding to initial maximum horizontal extent of $x_{\max,0} = 1200$~m and initial maximum vertical extent of about 33 m, as reported in \cite{shadab2025vertically}. The computational domain spans $x \in [0, 2.5\,x_{\max,0}]$~m horizontally and $z \in [0, 40]$~m vertically and is discretized uniformly into an $80 \times 40$ grid, giving a grid cell size of approximately 38 m in the horizontal direction and $\sim$1 m in the vertical direction. No-flow boundary conditions are imposed along all domain boundaries. For this experiment, we neglect heat conduction in order to verify and validate the numerical model's ability to simulate the advective flow of liquid in variably saturated firn. The refreezing here thus occurs due to advection of water into cold ice rather than due to conductive heat loss.

As the liquid water spreads horizontally, freezing occurs at the boundary between the firn aquifer at 0$\,^\circ$C and the cold region. The HydroFirn model shows excellent agreement with the analytic solutions. The quantitative estimates of dimensionless aquifer height $h_{max}$, horizontal extent $x_{max}$, and liquid water volume $V$ of the aquifer also show excellent agreement with the analytic solution (Figures~\ref{fig3:firn-aquifer}\textit{j-l}). As the aquifer expands, the maximum height decreases with time, the horizontal extent increases, and the liquid water volume decreases with time. The liquid-water volume is reduced by about 2\% at $t=10$ years due to heat advection caused meltwater freezing (Figure~\ref{fig3:firn-aquifer}\textit{l}) as the aquifer invades cold regions.

Applying HydroFirn in both the one-dimensional infiltration and two-dimensional aquifer migration test cases yields results that agree very well with the analytic solutions. These results give us confidence that the continuum model formulation and its numerical implementation are accurate. 

\section{ Two-dimensional meltwater infiltration in heterogeneous firn At DYE-2 site in SouthWest Greenland} \label{sec5:Dye2_study}
\begin{figure}
    \centering
    \includegraphics[width=0.8\linewidth]{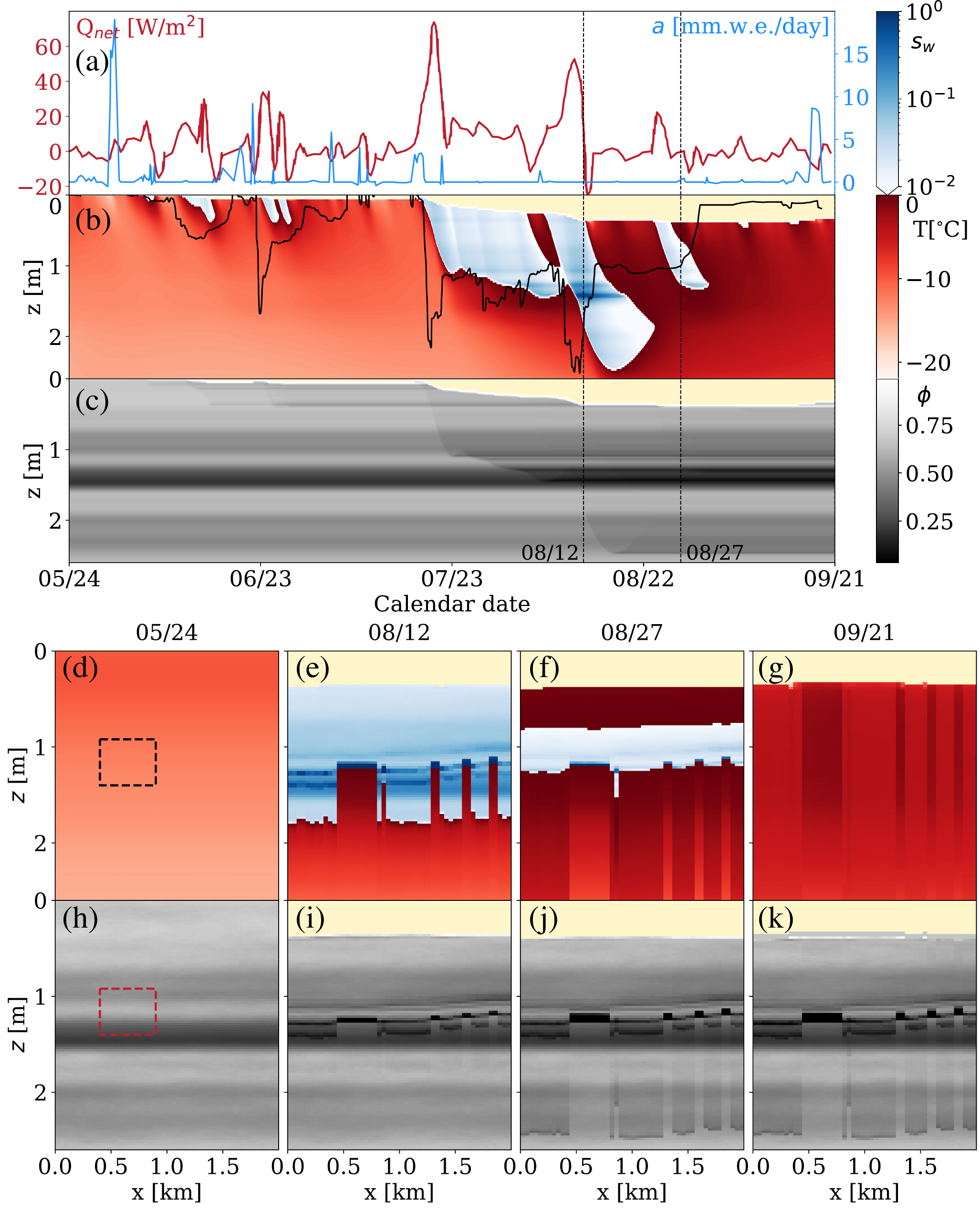}
    \caption{Modeled summer melting, percolation, and refreezing at DYE-2 site during summer of 2016  \citep[for more information about the data, see][]{vandecrux2020,samimi2020meltwater}: (a) Applied accumulation rate \textit{a} and applied surface thermal flux $Q_{net}$ leading to one-dimensional evolution of (b) combined saturation $s_w$ and firn temperature $T$ and (c) porosity $\phi$. Water percolation depths from upward-facing ground-penetrating radar measurements (upGPR, black line {in panel \textit{b}}) \citep{heilig2018seasonal} are generally consistent with the predicted saturation. Lower panels show the multidimensional evolution of heterogeneous firn with the same applied forcing (panel \textit{a}) leading to modeled (d,e,f,g) saturation $s_w$ and temperature T [$^\circ$C], and (h,i,j,k) porosity $\phi$ on (d,h) 05/24, (e,i) 08/12, (f,j) 08/27, and (g,k) 09/21. {The colorbars are shared among panels \textit{b-k}.} Heterogeneity is introduced in the porosity profile by overlapping the 1D distribution with a correlated random field with amplitude $\mathcal{A}$ = 0.05 and x-directional correlation length of 4000 m. The figure shows irregular percolation depths of meltwater and formation of low-porosity layers. There is no percolation below newly developed impermeable ice layers at 4 locations about a meter deep (panel \textit{f}) formed due to refreezing of higher localized melt (panel \textit{k}). The largest impermeable ice layer (>200 m length) is formed within the box made up of dashed lines in panels \textit{d,h}. The saturation contour is shown at a log scale to improve the contrast. Supplementary video S2 shows the corresponding evolution of firn corresponding to Figures~\ref{fig4:DYE-2_heterogeneous}\textit{d-k} and Figure~\ref{fig:zoomed-DYE2-corr-rnd-field}.}
    \label{fig4:DYE-2_heterogeneous}
\end{figure}

Finally, we simulate a field-scale case using observations from the DYE-2 site during the summer of 2016 \citep{samimi2020meltwater,vandecrux2020,heilig2018seasonal}. The one-dimensional domain of $z\in [0,5]$ m is divided into 200 cells with a grid resolution of 2.5 cm. The model is initialized using depth-dependent porosity and temperature profiles derived from field measurements (see \cite{shadab2024mechanism} for more information), and the firn surface is allowed to evolve under prescribed heat flux and snow accumulation boundary conditions constrained by observations from \cite{samimi2021time} (Figure~\ref{fig4:DYE-2_heterogeneous}\emph{a}). A prominent pre-existing discontinuity in porosity (precursor ice layer) exists at a depth of 1.2-1.5~m that is still permeable (firn porosity $\phi\sim 0.19$, density $\sim740$ kg/m$^3$). {In this paper, we use the adjectives ``new'' and ``pre-existing'' to highlight the temporal existence of ice layers, whereas we refer to the sharp discontinuities in porosities as ice layers only when the porosity goes below the cut-off porosity ($\phi<\phi_c$); otherwise, we refer to it as a precursor ice layer.} We further investigate how meltwater percolation and refreezing might vary spatially by synthesizing a 2D firn porosity field. To do so, we superimpose a 2D correlated random field for the firn porosity or permeability on top of 1D measured data (Appendix~\ref{sec:correlation}). Stratigraphic observations indicate that spatial correlations in firn arise from processes such as internal layering, spatiotemporal variability, and fluctuations in snow accumulation, with correlation lengths spanning a wide range of scales \cite[e.g.,][]{laepple2016layering,xu2023polar}.

\begin{figure}
    \centering
    \includegraphics[width=0.9\linewidth]{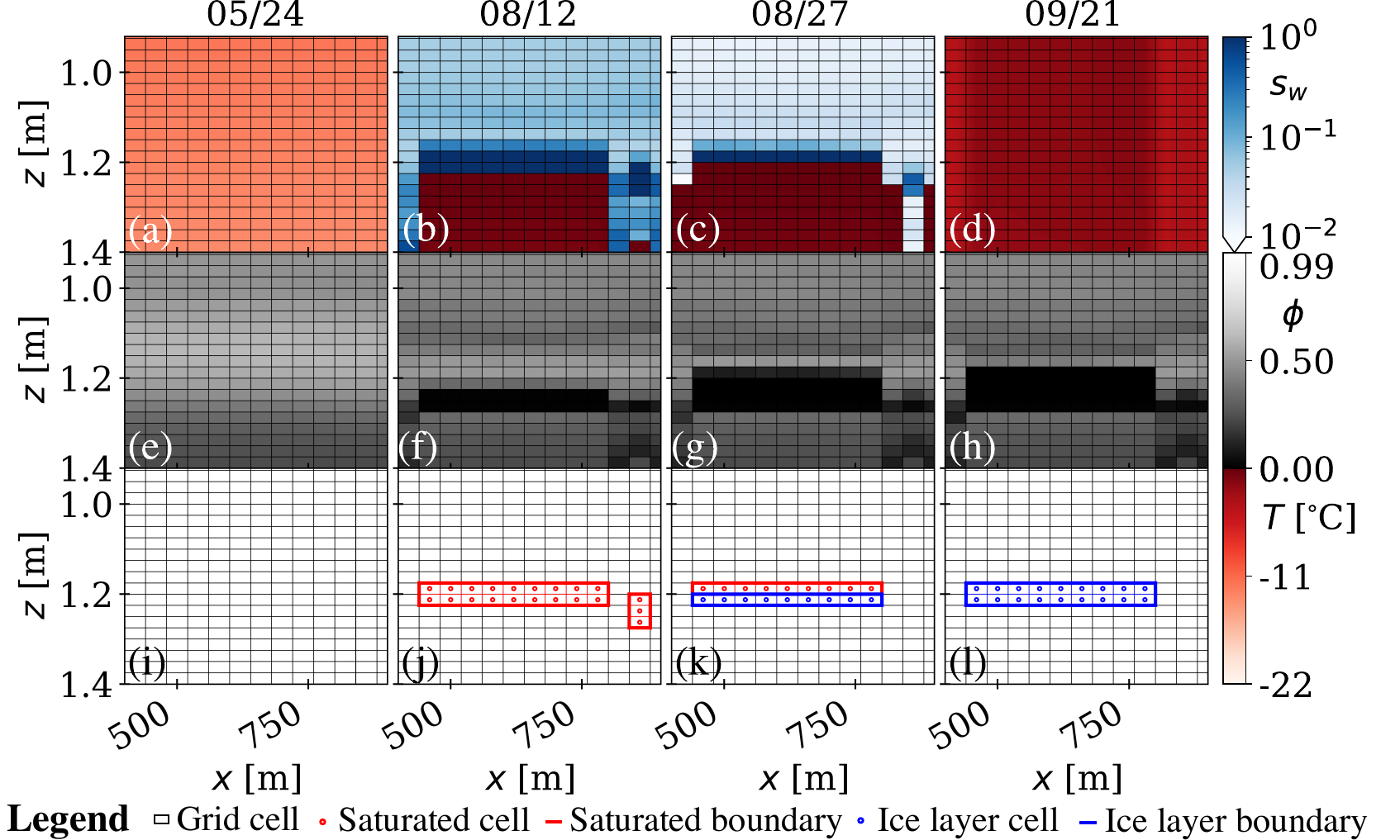}
    \caption{Zoomed figure showing summer melting 2016 at DYE-2 site corresponding to the dashed boxes in Figures~\ref{fig4:DYE-2_heterogeneous}\textit{d},\textit{h} showing modeled infiltration, perched water table formation, and its freezing to form impermeable ice layer formation. Panels show gridded spatial evolution of (a,b,c,d) saturation $s_w$ and temperature T [$^\circ$C], (e,f,g,h) porosity $\phi$, (i,j,k,l) cell classifications  for laterally heterogeneous firn on (a,e,i) 05/24, (b,f,j) 08/12, (c,g,k) 08/27, and (d,h,l) 09/21. {The colorbars are shared among panels \textit{a-h}.} The figure shows impermeable ice layer formation, saturated region formation and their interaction and evolution.}
    \label{fig:zoomed-DYE2-corr-rnd-field}
\end{figure}

{The net heat flux $Q_{\mathrm{net}}$ is calculated from surface energy balance monitored by the automatic weather station (AWS) \cite[see][for more information]{samimi2020meltwater} near the site. The timeseries of $Q_{\mathrm{net}}$ exhibits two intense melt events on 18-23 July and 7-12 August 2016, separated by a period of weaker but sustained positive forcing} ($\sim$10~W/m$^{2}$, Figure~\ref{fig4:DYE-2_heterogeneous}\textit{a}). This forcing leads to progressive firn warming and increasingly deep meltwater penetration (Figure~\ref{fig4:DYE-2_heterogeneous}\emph{b}). The presence of meltwater means the local temperature is at the melting temperature $0^\circ$C, and therefore it is replaced with the water saturation $s_w$. During the intervening period between the two intense events (23 July - 6 August 2016), reduced melt supply slows the wetting front's downward propagation and promotes refreezing above the pre-existing precursor ice layer, further decreasing its porosity; however, it remains permeable. During the 7-12 August melt event, meltwater penetrates beyond the pre-existing precursor ice layer as the local meltwater flux stays below its saturated hydraulic conductivity. This meltwater then refreezes to form a new precursor ice layer at $\sim$2.4~m depth that is still permeable (Figures~\ref{fig4:DYE-2_heterogeneous}\emph{b},\emph{c}). Here we refer to a region with a sharp reduction in porosity with respect to background porosity as a precursor ice layer that is still permeable. But once the pore space is closed off ($\rho > 830$ kg/m$^3$), we refer to such regions as (impermeable) ice layers. 
The modeled maximum penetration depth of the liquid water, where a  new precursor ice layer has formed, agrees with meltwater penetration depths inferred from upward-facing ground-penetrating radar during the 7-12 August melt event {\citep[black line in Figure~\ref{fig4:DYE-2_heterogeneous}\emph{b};][]{heilig2018seasonal}}. The radar data, however, suggest that the wetting front propagates downward faster at the onset of melting than the model simulates. {Increasing the permeability alone cannot resolve this discrepancy because it also overpredicts the penetration depth. Instead, it likely reflects processes not included in the present model, particularly local thermal disequilibrium \citep{rees2008local,moure2023thermodynamic} and preferential flow \citep{nimmo2012preferential,nimmo2021processes}, both of which can accelerate and deepen meltwater infiltration.}

We next investigate how meltwater percolation and refreezing might vary spatially using a synthetic 2D porosity field based on the field data. We now consider a 2D domain of $x\in [0,2000] \text{ m} \times z\in[0,5]$ m divided into 50$\times$200 cells, providing a resolution of 40 m and 2.5 cm in the $x$ and $z$ directions, respectively. To create this synthetic layer on top of a measured vertical variation in porosity, we impose a correlated random field with a correlation length in the x-direction $\theta_x = 4000$ m and an amplitude of $\mathcal{A}$ = 0.05. Detailed methodology along with other parameters is provided in Appendix~\ref{sec:correlation}. Figure~\ref{fig4:DYE-2_heterogeneous}\textit{h} shows the resulting porosity field along a 2 km lateral transect at the initial date of 05/24/2016, which features the lateral heterogeneities due to a correlated random field $\phi_{corr}(x,z)$ superimposed on top of the vertical variation in porosity $\phi_{meas}(z)$.

Our results show that lateral heterogeneity strongly affects surface melting, meltwater migration, and final porosity distribution (stratigraphy). In the beginning, the temperature profile shows minimal changes but the porosity profile shows lateral variations that are less significant due to correlated random field (Figures~\ref{fig4:DYE-2_heterogeneous}\textit{d,h}, zoomed version in Figures~\ref{fig:zoomed-DYE2-corr-rnd-field}\textit{a,e}). On August 7, 2016, there are two wet regions in firn (Figure~\ref{fig4:DYE-2_heterogeneous}\textit{b}): the first is close to the surface and results from the most recent melting that started around August 7, 2016, and the second is at a depth of 1-1.4 m, resulting from the July 18-20 melt event. When the melting is complete on August 12, perched water tables (saturated regions) have formed on the pre-existing precursor ice layer at around 1.2 m depth (Figures~\ref{fig4:DYE-2_heterogeneous}\textit{e}, zoomed Figure~\ref{fig:zoomed-DYE2-corr-rnd-field}\textit{b}). Figure~ \ref{fig:zoomed-DYE2-corr-rnd-field}\textit{j} shows the corresponding numerical grid, saturated cells $\Omega$ (red circles), and saturated region boundary $\partial \Omega_s$ (red lines). {It is clear that the depth of meltwater percolation along the 2 km transect is not constant but depends on the porosity distribution that exists near the surface (Figures~\ref{fig4:DYE-2_heterogeneous}\textit{d-k}, zoomed Figure~\ref{fig:zoomed-DYE2-corr-rnd-field}). At the surface, porosity controls both the thermal conductivity and the thermal storativity, thereby determining the thermal penetration depth. Consequently, for a given thermal flux, higher porosity increases the near-surface snow temperature (the flux required to cause melting scales as $(1-\phi)^{0.44}$; see \cite{shadab2024mechanism}). Below the surface, the local porosity controls the speed of melt percolation.} At four locations, the pores have completely closed off due to refreezing and formed an impermeable ice layer (Figures~\ref{fig4:DYE-2_heterogeneous}\textit{k}). Not all of the saturated regions freeze to form impermeable ice layers (Figures~\ref{fig:zoomed-DYE2-corr-rnd-field}\textit{j}-\textit{l}). The temperatures also vary spatially in regions where melt can or cannot percolate due to the absence or presence of ice layers (Figure~\ref{fig4:DYE-2_heterogeneous}\textit{f}). This happens as heat conduction is typically a slower heat transport process than heat advection by liquid water. {In summary, lateral heterogeneity that is initially difficult to discern may strongly affect the dynamics and final distribution of meltwater, porosity, temperature, and ice-layer formation within firn.}

\begin{figure}
    \centering
    \includegraphics[width=0.75\linewidth]{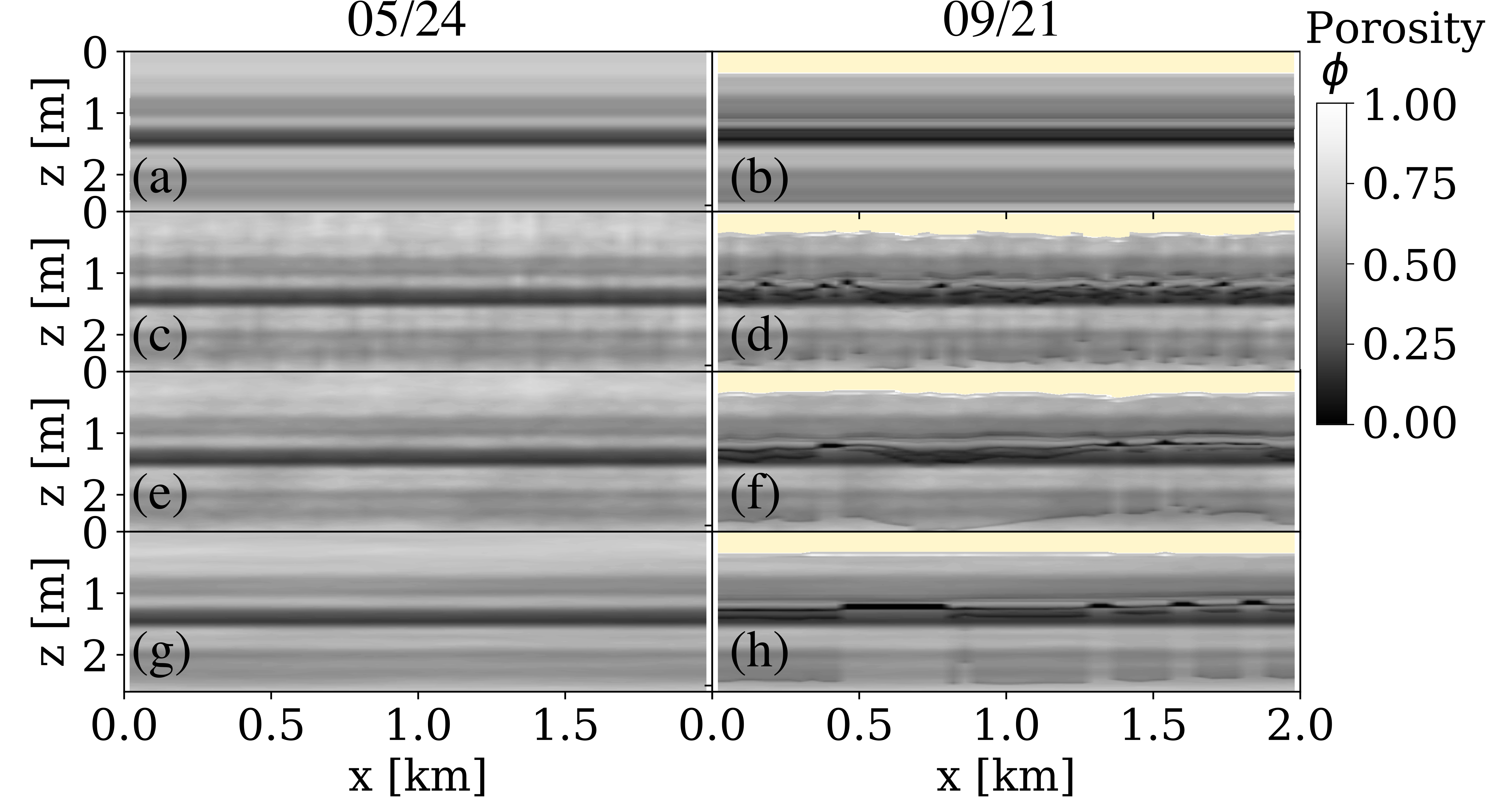}
    \caption{Effect of correlation length on infiltration in laterally heterogeneous firn for an amplitude of 0.05. Porosity contours for (a,b) laterally homogeneous case and for heterogeneous case with correlation lengths of (c,d) 40 m, (e,f) 400 m, and (g,h) 4000 m with former being the initial state on 05/24 and latter being the final state on 09/21.}
    \label{fig5:eff-of-corr-length}
\end{figure}

We ran model experiments using the same surface forcings but different initial porosity fields to test how the correlation length and amplitude affect our results. The correlation length in the x-direction (Figure~\ref{fig5:eff-of-corr-length}) controls how far the porosity distribution is related, reducing the {lateral variability}. Figure~\ref{fig5:eff-of-corr-length}\textit{a} shows the initial state on May 24, 2016 for a laterally homogeneous transect. Increasing the correlation length from 40 m (Figure~\ref{fig5:eff-of-corr-length}\textit{c}) to 400 m (Figure~\ref{fig5:eff-of-corr-length}\textit{e}) and to 4000 m (Figure~\ref{fig5:eff-of-corr-length}\textit{g}, also shown earlier in Figures~\ref{fig4:DYE-2_heterogeneous}\textit{d-k} and Figure~\ref{fig:zoomed-DYE2-corr-rnd-field}) causes the transect to become increasingly similar to a laterally homogeneous transect. A shorter correlation length leads to more lateral variability in the initial distribution. The corresponding final porosity distributions show that the melting has become {increasingly laterally uniform} with increased correlation length (Figures~\ref{fig5:eff-of-corr-length}\textit{b},\textit{d},\textit{f},\textit{h}). For shorter correlation lengths, the final distribution shows variability in surface melting, depth of percolation, and a higher number of small, disconnected ice layers right above the pre-existing ice layer. Additionally, length of the impermeable ice layer over the pre-existing ice layer has increased, but the number of small, disconnected impermeable ice layers has decreased with the correlation length, even though these layers have become {increasingly laterally uniform}.

\begin{figure}
    \centering
    \includegraphics[width=0.75\linewidth]{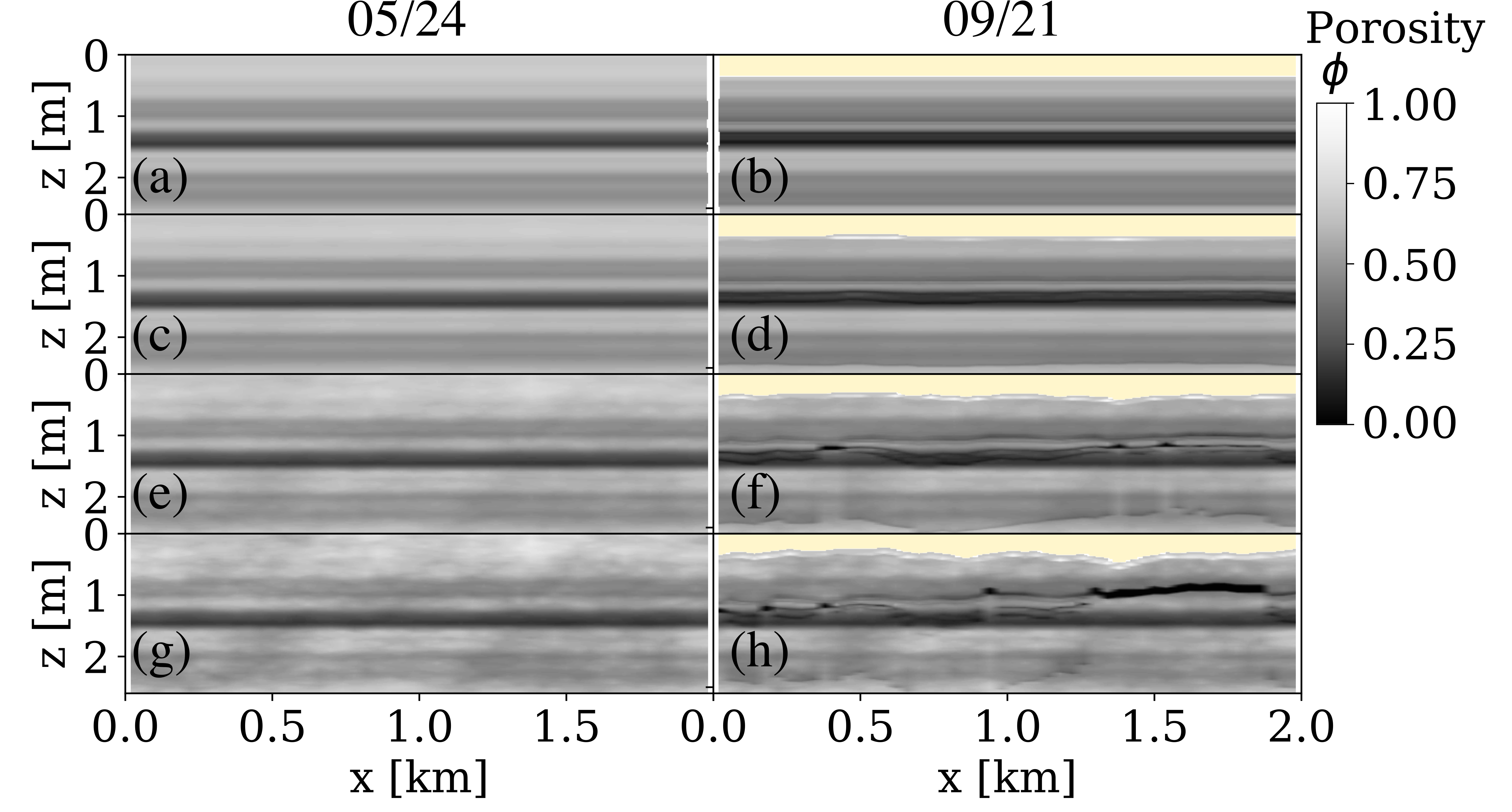}
    \caption{Effect of amplitude on infiltration in laterally heterogeneous firn for a correlation length = 400m. Porosity contours for (a,b) laterally homogeneous case and for heterogeneous case with Amplitude $\mathcal{A}$ of (c,d) 0.01, (e,f) 0.05, and (g,h) 0.1 with former being the initial state on 05/24 and latter being the final state on 09/21.}
    \label{fig6:effect_of_amplitude}
\end{figure}

The amplitude of the correlated random field controls its strength (Figure~\ref{fig6:effect_of_amplitude}), i.e., how high the highs are and how low the lows are in porosity with respect to the local value. Increasing the amplitude leads to a more prominent, laterally heterogeneous porosity distribution (Figures~\ref{fig6:effect_of_amplitude}\textit{a},\textit{c},\textit{e},\textit{g}). Unsurprisingly, the increasing amplitude of porosity variation affects the final distribution of the porosity on September 21, 2016 (Figures~\ref{fig6:effect_of_amplitude}\textit{b},\textit{d},\textit{f},\textit{h}). As the amplitude increases, the variability in the final porosity also increases relative to the homogeneous scenario (Figures~\ref{fig6:effect_of_amplitude}\textit{a},\textit{b}). For example, with $\mathcal{A}$ = 0.01, there is very little difference between the final porosity and that of the laterally homogeneous 1D scenario. On the other hand, in the highest amplitude experiment ($\mathcal{A}$ = 0.1), there is a prominent impermeable ice layer of varying thickness at about 1.2 m depth, spanning $x=$1.3 km to 1.9 km, right above the pre-existing precursor ice layer (Figures~\ref{fig6:effect_of_amplitude}\textit{h}).

\begin{figure}
    \centering
    \includegraphics[width=0.45\linewidth]{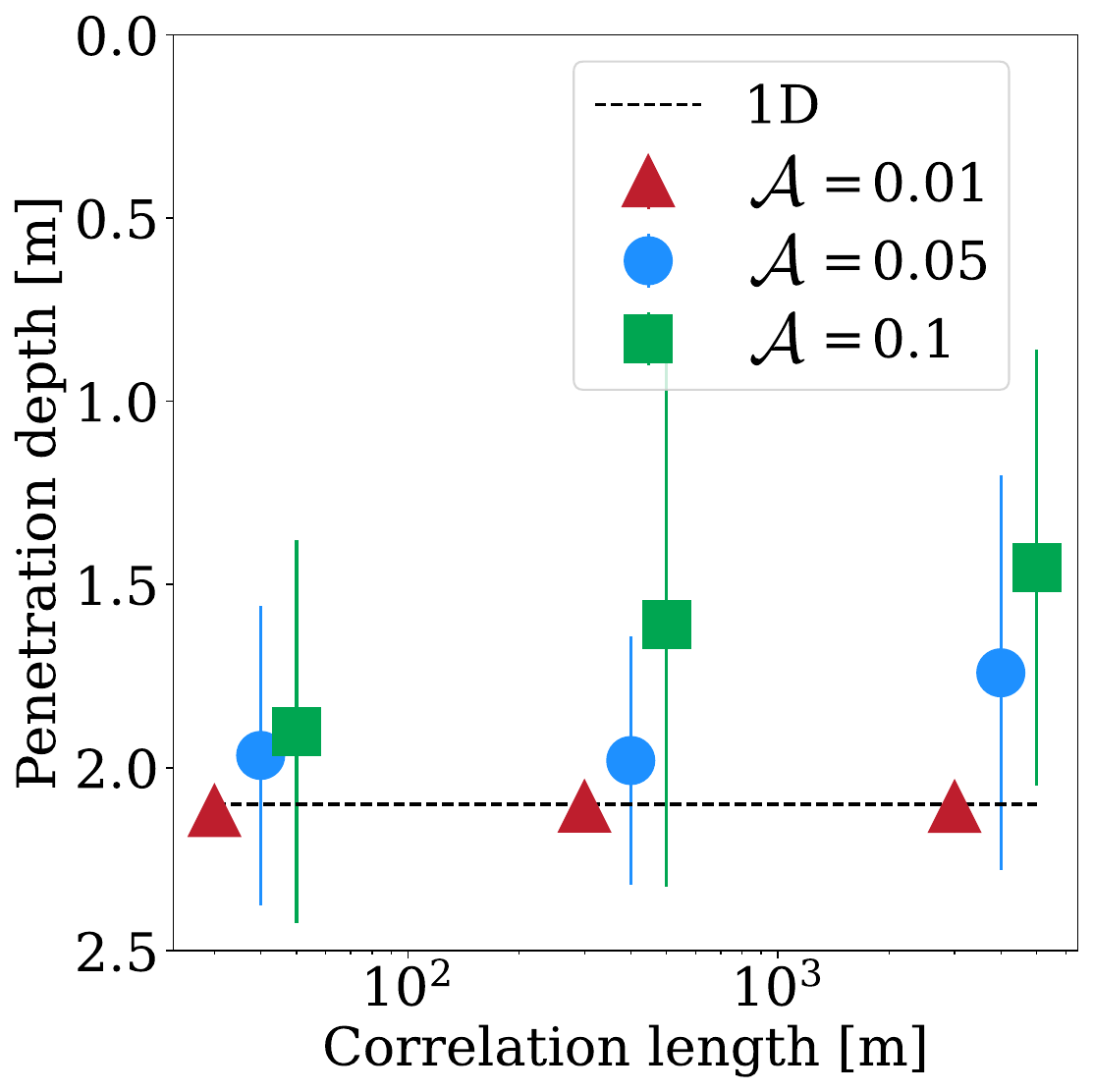}
    \caption{Dependence of penetration depth on correlation length and amplitude of the correlated, random field. Penetration depth is defined as the deepest point of melt percolation where a new precursor ice layer may form due to refreezing. The markers are offset in x-direction for clarity. The correlation length for each cluster is given by the corresponding blue marker {and the vertical bars show one standard deviation.}}
    \label{fig7:summary-of-corr-rnd-field}
\end{figure}

Lateral heterogeneity affects the deepest vertical location where melt has percolated and refrozen to form a precursor ice layer, referred to as the penetration depth. An increased correlation length leads to more uniform penetration depths across the transect {(Figures~\ref{fig5:eff-of-corr-length}\textit{b},\textit{d},\textit{f},\textit{h})}. However, the mean penetration depth becomes shallower as larger, more prominent impermeable ice layers become shallower ($\mathcal{A}$ = 0.05 and 0.1 in Figure~\ref{fig7:summary-of-corr-rnd-field}). The percolation depth at amplitude $\mathcal{A}$ = 0.01 (i.e., low amplitude) is essentially identical to the laterally homogeneous case because the lateral heterogeneity in porosity is not significant. However, increasing the amplitude further causes a shallower mean penetration depth and increases the standard deviation as the meltwater starts to form perched water tables over laterally heterogeneous pre-existing precursor ice layers.

\section{ Discussion and Conclusions}\label{sec6:conclusion}

This study introduces \textit{HydroFirn}, a large-scale, multidimensional, multiphase thermo-hydrologic model for firn that resolves coupled meltwater transport, heat transport, and phase change with realistic climatic forcing that includes net thermal flux and snow accumulation. The model is designed to represent the transition from gravity-driven unsaturated flow to hydraulic pressure-driven saturated flow, including the emergence and evolution of firn aquifers, perched water tables, and impermeable ice layers in a computationally efficient framework suitable for large spatial domains.

HydroFirn solves conservation laws for composition and enthalpy with constitutive relations that differentiate between unsaturated and saturated regimes. The numerical implementation employs the conditionally implicit pressure and explicit enthalpy and composition (CIMPEC) algorithm, which only solves the additional pressure (head) equation on saturated cells, enabling efficient simulations at a large scale.
The model reproduces analytic solutions for (i) one-dimensional infiltration through {laterally uniform} firn leading to formation of a perched water table and (ii) two-dimensional lateral expansion of a cold firn aquifer over an impermeable base. These benchmarks test coupled thermodynamics, phase change, and saturated-unsaturated region transitions, and HydroFirn shows excellent agreement in both 1D and 2D.

\begin{figure}
    \centering
    \includegraphics[width=\linewidth]{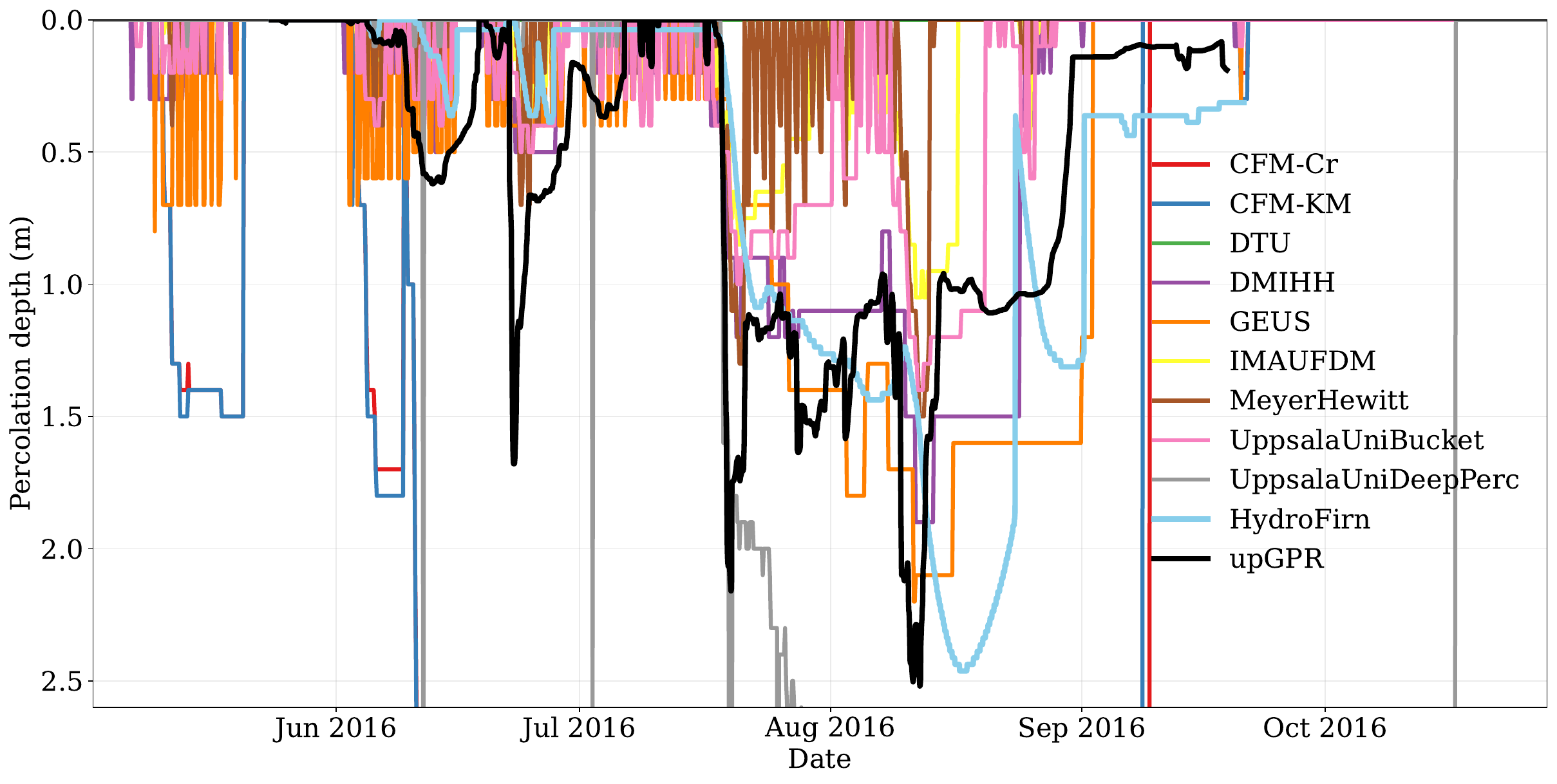}
\caption{Simulated and observed meltwater percolation depth at DYE-2 during the 2016 melt season corresponding to Figures~\ref{fig4:DYE-2_heterogeneous}\textit{a-c}. Colored lines show the instantaneous deepest depth containing liquid water in the nine RetMIP firn models in addition to HydroFirn model for one-dimensional simulation. See \cite{vandecrux2020} for more information about the firn models. The black line shows the percolation depth derived from upGPR observations from \cite{heilig2018seasonal}.}
    \label{fig:DYE-2_percolation_depth}
\end{figure}

When forced with observed surface heat flux and accumulation at DYE-2 site, the model captures progressive firn warming, deepening meltwater penetration, and the formation of a new low-porosity layer at depths consistent with radar-inferred meltwater penetration during the August 2016 melt event. {Figure~\ref{fig:DYE-2_percolation_depth} compares the deepest point of the melt presence, referred to as percolation depth, from the present model as well as nine firn models in the Retention Model Intercomparison Project (RetMIP) \citep[see][for more information on the nine firn models]{vandecrux2020} and field measurements from upward-facing ground-penetrating radar (upGPR) \citep{heilig2018seasonal}. All nine RetMIP models exhibit substantial differences in both the timing and depth of meltwater penetration. Most firn models simulate predominantly shallow infiltration within the upper approximately 2~m, whereas others (e.g., CFM-Cr, CFM-KM, UppsalaUniDeepPerc) predict substantially deeper percolation that extends beyond the displayed 2.6~m range. Using the set up of \cite{shadab2024mechanism}, HydroFirn reproduces the observed seasonal progression from shallow infiltration to a maximum percolation depth of approximately 2.5~m in August, followed by a retreat of the liquid-water front toward the surface. The timing and magnitude of this maximum are broadly consistent with the upGPR observations, although differences remain in the short-term evolution and persistence of the simulated wetting front. The large intermodel spread demonstrates the sensitivity of simulated percolation to the representation of water transport and retention in firn.}

Introducing lateral heterogeneity along a transect via correlated random fields in a 2D model experiment demonstrates that lateral structure in porosity/permeability can alter the depth and spatial pattern of meltwater percolation and the resulting ice-layer geometry. In particular, stronger heterogeneity (larger amplitude) increases variability in penetration depth and promotes localized impermeable-layer formation above pre-existing stratigraphic discontinuities. This 2D experiment was an idealized simulation to test how lateral heterogeneity may affect meltwater percolation and refreezing. 
{However, field measurements of the amplitude and spatial scale of variability in firn properties such as density are limited \citep{scott2006ground,parry2007investigations,brown2011high}.} {The assumption of representative elemental volume is crucial for the choice of grid size since it may lead to smearing of strong discontinuities within the grid cell \citep{vandecrux2020}. This is both an issue with collecting density measurements \citep{rennermalm2022shallow} as well as while simulating using Eulerian models \citep{vandecrux2020}. For all DYE-2 simulations in Section 5, the resolution was chosen to be 2.5 cm by validating against the melt percolation depth from the field data, that comes from three sources (time domain reflectometers \citep{samimi2021time}, upGPR \citep{heilig2018seasonal} and passive remote sensing \citep{colliander2022ice}).} {Our results underscore the need for additional field measurements spanning spatial scales in order to better understand the relationship between (common) point scale measurements and larger-scale (e.g., 10-100 km) patterns of firn properties and processes that may, e.g., be simulated in earth system models}.



Several processes remain outside the scope of the present model and motivate future developments. First, the formulation neglects capillary effects, which may influence wetting-front structure and hysteresis in certain regimes. Extending the framework to include capillary pressure would broaden its applicability. Second, the evolution of firn microstructure (grain size, anisotropy, and evolving permeability relations) and firn compaction are not considered here. Coupling HydroFirn to a dynamic compaction and microstructure module would enable longer-term simulations with evolving stratigraphy. Third, preferential flow and channelized transport are not represented explicitly. Incorporating subgrid or dual-domain (matrix and preferential flow) parameterizations would allow for the assessment of their influence on ice-layer formation and runoff. {Fourth, the model neglects water vapor transport and vapor-ice phase exchange. Although these processes are expected to have a limited influence on short-term meltwater percolation and refreezing, they play an important role in firn metamorphism, grain growth, and the long-term evolution of firn microstructure \citep{albert2004extreme,firn2024firn,mcdowell2023firn}. Incorporating water vapor transport would therefore broaden the applicability of HydroFirn to studies of firn evolution and air-firn interactions \citep{albert2004extreme,firn2024firn,mcdowell2023firn}. Finally, extending the current multidimensional implementation to fully three-dimensional domains and integrating spatially distributed forcing from remote sensing and regional climate models will enable ice-sheet-scale investigations of lateral meltwater routing and ice-layer evolution.}

HydroFirn represents an important step forward in firn hydrology modeling. By explicitly resolving multidimensional routing and perching processes, HydroFirn provides physically based constraints on where and when meltwater is stored, refreezes, or contributes to runoff. These constraints are directly relevant for interpreting wet firn hydrology and densification, for reducing uncertainty in converting altimetric elevation change to mass change, and for improving estimates of freshwater export under a warming climate. 
{From a liquid water balance (LWB) perspective \citep{steger2017modelled}, HydroFirn model can help estimate surface mass fluxes of rainfall, evaporation, and meltwater, internal refreezing, and subsurface runoff (including the lateral component) except the surface runoff where the contribution of the surface runoff might be significant.} {Furthermore, the model can help assess the location-dependent length scales of horizontal flow which will help parametrize subsurface runoff in lateral direction in one-dimensional models which is an excellent direction of future research.} Overall, HydroFirn provides a verified and computationally efficient platform for investigating multidimensional firn hydrology, including saturated regions and impermeable ice layers, and for connecting local processes to large-scale consequences for ice-sheet mass balance.

\section*{ Supplementary material}
Supplementary video S1 shows the modeled expansion of the firn aquifer in an otherwise $-30^\circ$C firn given by HydroFirn model in the absence of heat conduction (contour plots or solid blue lines) compared against semi-analytical solutions (red dashed lines) corresponding to Figure~\ref{fig3:firn-aquifer}. Furthermore, Supplementary video S2 shows the numerical solutions for the infiltration in the heterogeneous layer at DYE-2 during the summer of 2016, corresponding to Figure~\ref{fig4:DYE-2_heterogeneous}.  The video shows combined saturation $s_w$, firn temperature $T$, and porosity $\phi$ for the one dimensional case and its corresponding two dimensional, laterally heterogeneous counterpart for an amplitude of 0.05 and a horizontal correlation length of 4 km.

\section*{ Acknowledgments and Funding} 
M.A.S. was supported through Princeton University's Future Faculty in Physical Sciences Postdoctoral Fellowship. C.M.S. was supported by NASA Grant 80NSSC25K7216. {The code and data used to generate figures of this paper are available on Github (link: \url{https://github.com/mashadab/HydroFirn}) and archived on Zenodo \citep{shadab2026hydrofirn} for reproducibility.} The authors acknowledge the initial discussions with Cyril Grima and Anja Rutishauser that motivated this work.

\bibliography{igsrefs,bibliography_stack1,bibliography_stack2}   
\bibliographystyle{igs}  

\appendix

\section{ Generation of correlated random fields}\label{sec:correlation}

In Section~\ref{sec5:Dye2_study}, spatial heterogeneity in both porosity and permeability is introduced using correlated random fields. This is done by multiplying the measured vertical variation with the correlated random field. The spatial variability is assumed to arise from elliptically shaped structures, leading to a transversely anisotropic exponential correlation model. Following \cite{zhu2013characterizing} and \cite{shadab2024hyperbolic}, the correlation function $\varrho$ is expressed as

\begin{align}\label{eq:correlation-fxn}
    \varrho = \exp\!\left(-2\sqrt{\frac{\Delta \mathcal{X}^2}{\theta_x^2} + \frac{\Delta \mathcal{Z}^2}{\theta_z^2}}\right),
\end{align}

where $\Delta \mathcal{X}$ and $\Delta \mathcal{Z}$ denote the horizontal and vertical separation distances between two spatial locations. The parameters $\theta_x$ and $\theta_z$ represent the characteristic correlation lengths in the horizontal and vertical directions, respectively.

To generate realizations of the correlated random field, we employ a matrix decomposition approach. A discrete set of $N$ spatial locations is first defined, at which the random field is sampled. Using the correlation function in Equation~\eqref{eq:correlation-fxn}, an $N \times N$ covariance matrix $\textit{\textbf{C}}$ is constructed, with each entry representing the covariance between a pair of sampling points. An exact, though computationally demanding, Cholesky decomposition \citep{trefethen1997numerical} is then applied to factorize the covariance matrix into lower and upper triangular matrices,

\begin{align*}
    \textit{\textbf{C}} = \textit{\textbf{L}}\,\textit{\textbf{L}}^{T}.
\end{align*}

\begin{figure}
    \centering
    \includegraphics[width=0.8\linewidth]{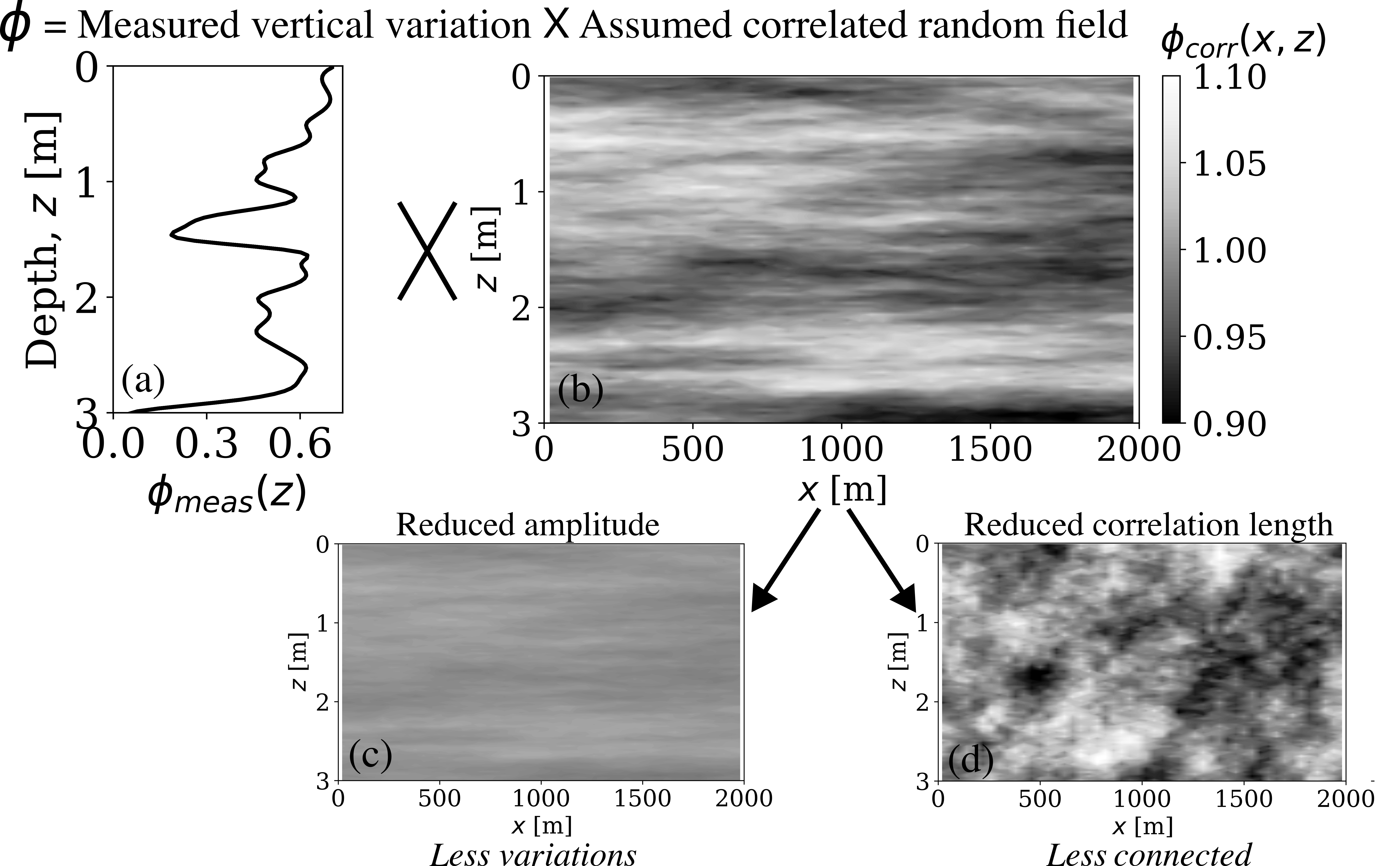}
    \caption{Enforcing lateral heterogeneity: (a) observed one-dimensional porosity $\phi_{meas}(z)$, which is multiplied with (b) a normalized, correlated random field of amplitude $\phi_{corr}(x,z)$, $\mathcal{A}$=0.05 and correlation length=4000 m. Normalized correlated random field $\phi_{corr}$ for (b) $\mathcal{A}$=0.01 and correlated length 4000 m and (c) $\mathcal{A}$=0.05 and correlated length of 400 m. The field $\phi_{corr}$ depends on amplitude ($\mathcal{A}$) and  correlation length.}
    \label{figS1:corr-random-field}
\end{figure}

A vector $\textit{\textbf{X}}$ of length $N$, consisting of independent standard normal random variables, is generated next. The correlated random field $\textit{\textbf{Y}}$ is obtained by multiplying $\textit{\textbf{X}}$ with the Cholesky factor and adding the mean vector $\boldsymbol{\mu}$,

\begin{align}
    \textit{\textbf{Y}} = \mathcal{A} \cdot \textit{\textbf{L}}\,\textit{\textbf{X}} + \boldsymbol{\mu}.
\end{align}
where $\mathcal{A}$ being the amplitude of the correlated field. Because permeability typically varies over several orders of magnitude, the absolute permeability field is computed using a logarithmic transformation, $\textbf{k} = 10^{\textit{\textbf{Y}}}$, yielding an $N \times 1$ vector of permeability values associated with individual grid cells. The corresponding measured porosity field $\phi(\textbf{x})$ is then derived by multiplying the local porosity $\phi_{meas}(\textbf{z})$ by $\phi_{corr}=\textbf{k}^{1/m}$ such that

\[
\phi(\textbf{x}) = \phi_{meas}(z)\phi_{corr}(\textbf{x}).
\]
The one-dimensional porosity variation (Figure~\ref{fig4:DYE-2_heterogeneous}\textit{b}) using this approach can be converted to the two-dimensional field (Figure~\ref{fig4:DYE-2_heterogeneous}\textit{g}) for $\theta_z=1$ m, $\theta_x = 4000$ m, $\mathcal{A}$ = 0.05, and $\boldsymbol{\mu}=\textbf{0}$. Although $\phi_{corr}$ is a correlated random field, the function $\phi(\textbf{x})$ may not, however, be completely correlated due to one-dimensional variation $\phi_{meas}(z)$. The rest of the plots in Section~\ref{sec5:Dye2_study} are plotted by changing either the x-correlation length $\theta_x$ or the amplitude $\mathcal{A}$. An example workflow for the generation of the {laterally} heterogeneous porosity field is given in Figure~\ref{figS1:corr-random-field}.

\end{document}